\documentclass{emulateapj}
\slugcomment{Accepted to The Astrophysical Journal}
\pdfoutput=1
\usepackage{natbib}
\newcounter{species}

\begin{document}

\title{Simulations of MHD Instabilities in Intracluster Medium
Including Anisotropic Thermal Conduction}

\author{Tamara Bogdanovi\'c\altaffilmark{1}, Christopher
S. Reynolds\altaffilmark{1}, Steven A. Balbus\altaffilmark{2}, and Ian
J. Parrish\altaffilmark{3}}

\altaffiltext{1}{Department of Astronomy, University of Maryland,
College Park, MD 20742-2421, e-mail: {\tt tamarab, chris@astro.umd.edu}}

\altaffiltext{2}{\'Ecole Normale Sup\'erieure, Laboratoire de
Radioastronomie, 24 rue Lhomond, 75231 Paris CEDEX 05, France,
e-mail:{\tt steven.balbus@lra.ens.fr}}

\altaffiltext{3}{Astronomy Department and Theoretical Astrophysics
Center, 601 Campbell Hall, University of California, Berkeley, CA
94720, e-mail: {\tt iparrish@astro.berkeley.edu}}

\begin{abstract}
We perform a suite of simulations of cooling cores in clusters of
galaxies in order to investigate the effect of the recently discovered
heat flux buoyancy instability (HBI) on the evolution of cores.  Our
models follow the 3-dimensional magnetohydrodynamics (MHD) of cooling
cluster cores and capture the effects of anisotropic heat conduction
along the lines of magnetic field, but do not account for the
cosmological setting of clusters or the presence of AGN.  Our model
clusters can be divided into three groups according to their final
thermodynamical state: catastrophically collapsing cores, isothermal
cores, and an intermediate group whose final state is determined by
the initial configuration of magnetic field. Modeled cores that are
reminiscent of real cluster cores show evolution towards
thermal collapse on a time scale which is prolonged by a factor of
$\sim2-10\,$ compared with the zero-conduction cases. The
principal effect of the HBI is to re-orient field lines to be
perpendicular to the temperature gradient.  Once the field has been
wrapped up onto spherical surfaces surrounding the core, the core is
insulated from further conductive heating (with the effective thermal
conduction suppressed to less than $10^{-2}$ of the Spitzer value) and
proceeds to collapse.  We speculate that, in real clusters, the
central AGN and possibly mergers play the role of ``stirrers,''
periodically disrupting the azimuthal field structure and allowing
thermal conduction to sporadically heat the core.
\end{abstract}


\section{Introduction\label{S_intro}}

The temperature structure of the intracluster medium (ICM) in
central regions of galaxy clusters is bimodal.  The non-cooling core
clusters have isothermal ICM cores with low densities and hence long
radiative cooling times.  On the other hand, the central regions of
the ICM in cooling core clusters have radiative cooling times that can
be as short as $10^8$\,years.  Within the cooling radius (the location
where the radiative cooling time is comparable to the Hubble time and
hence the age of the cluster), the temperature decreases with
decreasing radius (Peterson \& Fabian 2006).  Given the short cooling
time, it is profoundly puzzling that the cores of these clusters have
not undergone a cooling catastrophe.  The current paradigm is that the
central active galactic nucleus (AGN) heats the ICM; this remains one
of the most direct arguments for ``AGN feedback''.  However, early
work (Binney \& Cowie 1981) highlighted the possible role that thermal
conduction may play in these clusters.  

Our understanding of the action of thermal conduction in atmospheres
such as the ICM is undergoing a revolution. Because the ICM plasma is
very dilute, thermal conduction will act in a very anisotropic manner,
occurring essentially unchecked along magnetic field lines but being
very strongly suppressed perpendicular to field lines.  This
elementary fact has profound implications for the dynamics and
structure of the ICM (or, indeed, any dilute plasma atmosphere).  As
shown by \citet{balbus00}, the anisotropy of conduction fundamentally
alters the Schwarzschild criterion for convection --- rather than
requiring an inverted {\it entropy} gradient, convection in such an
magnetohydrodynamic (MHD) atmosphere will occur whenever the {\it
temperature} gradient is inverted, due to the {\it magneto-thermal
instability}, \citep[MTI; see][for the first numerical studies of this
instability]{ps05,ps07}. \citet{psl08} studied the effect of the MTI
in the outer regions of clusters, where temperature decreases with
radius, and found that the temperature profile of the ICM can be
substantially modified on timescales of several billion years. The
instability drives field lines to become preferentially radial leading
to conduction at a high fraction of the Spitzer conductivity.

The ICM cores of cooling core galaxy clusters will be stable to the
MTI (since the temperature in such cores is increasing with radius).
However, \cite{quataert08} has discovered a related instability (the
{\it heat flux buoyancy instability}; HBI) which acts when the
temperature is increasing with radius.  Based on local simulations of
the HBI in 3D stratified atmosphere, \citet{pq08} found the HBI
induces MHD turbulence and can somewhat amplify the magnetic field in
the plasma. They also discovered that, in the plane-parallel geometry
that characterizes all local simulations, the instability saturates
when the lines of magnetic field are re-oriented in such a way as to
suppress the heat transport across the temperature gradient.

Guided by the results from the local simulations, \cite{br08}
suggested that MHD turbulence driven by the HBI may be important in
regulating the conduction of heat into ICM cores and could mediate the
stabilization of the cooling cores.  They hypothesized that the
presence of radiative cooling and the spherical (as opposed to planar)
geometry would prevent field line re-orientation from completely
insulating the core from the conductive heat flux.  They also
highlighted the fact that HBI driven turbulence would create a
convective heat flux that {\it removes} heat from the cool core (i.e.,
it acts as a cooling term in the energy equation).

In this paper, we present global models of cooling core clusters
in which we explore the role of heat conduction and the HBI on the
evolution of these cores. The results of our simulations suggest that
HBI alone cannot regulate and stabilize a cooling core.  We have
followed the non-linear evolution of the HBI in the inner $\sim {\rm
few} \times 100$~kpc in clusters, and found that it is ubiquitous and
rearranges the lines of magnetic field in such a way that they are
wrapped around the core. This results in dramatic suppression of the
heat conduction below the Spitzer value. Consequently, within context
of these simple models (which do not include AGN or realistic
cluster/dark matter dynamics) heat conduction can significantly delay
but cannot prevent the catastrophic core collapse in real clusters. In
\S~2 we briefly review the equations and time scales governing the
problem of heat conduction in cluster cores as well as the numerical
setup used in simulations. In \S~3 we present results from a parameter
space study of cooling core clusters and then focus on a specific
model of a core resembling that in the Perseus cluster. We discuss our
results, approximations, and observational consequences in \S~4, and
present our conclusions in \S~5.

\section{Simulations\label{S_methods}}

Using the 3-dimensional MHD code {\it Athena} \citep{stone08}, we have carried
out simulations of thermally conducting ICM cores which incorporate
the effects of anisotropic thermal conduction.  We run two classes of
simulations.  Firstly, we conduct a suite of simulations aimed at
mapping out the behavior of clusters as a function of position in the
2-dimensional parameter space $(t_{\rm cool}/t_{\rm dyn}, t_{\rm
cond}/t_{\rm dyn})$, where the cooling timescale $t_{\rm cool}$,
conduction timescale $t_{\rm cond}$, and dynamical timescale $t_{\rm
dyn}$ are defined in Section~\ref{S_timescales}.  These simulations
describe ``theorist clusters'' in the sense that no particular
physical parameters (e.g., density, temperature and timescales) are
implied, with only dimensionless ratios being relevant.  We shall
refer to these as our ``parameter space survey simulations''.
Secondly, we shall conduct simulations that are specifically tailored
to describe physically-realistic clusters.  We shall refer to these as
our ``physical clusters/simulations''.

We must note one peculiarity of these (or indeed any) simulations of
``idealized'' galaxy clusters.  In removing our model clusters from
their cosmological setting, the computational complexity of the
problem is reduced enormously.  Indeed, it is currently infeasible to
perform a cosmological simulation which resolves the small scale
physics relevant to our study.  However, it is important to realize
that galaxy clusters are dynamically young objects and, through the
accretion of subclusters and groups, are still in the process of
forming.  Any {\it ab initio} model for the thermodynamic state of the
ICM {\it must} acknowledge the cosmological setting.  Neglecting the
cosmological setting has two consequences for our work.  Firstly, real
systems will possess dynamics related to merging substructures that is
not captured in our treatment.  The effects of neglecting this
phenomenon will be discussed in Section~4.  Secondly, there is no
well-defined choice of ``physical initial conditions'' for our model
clusters.  We must be content with forming well defined initial
conditions that describe gross aspects of the systems under
consideration.  We describe our choice of initial conditions below.
The fact that our results agree well with those of Parrish et
al. (2009) who employ a rather different choice of initial condition
suggests a robustness to these choices.

\subsection{Equations\label{S_equations}}

The fundamental equations of the analysis are
\begin{eqnarray}
&&\frac{\partial\rho}{\partial t}  + \nabla\cdot(\rho {\bf v}) = 0 \\
&& \rho \frac{\partial{\bf v}}{\partial t} + \rho({\bf v}\cdot
\nabla){\bf v} = \frac{(\nabla\times{\bf B})\times{\bf B}}{4\pi}-
\nabla p + \rho{\bf g},\\
&& \frac{\partial{\bf B}}{\partial t} = \nabla\times({\bf v}\times{\bf
B}), \\
&& \frac{\partial e}{\partial t} + \nabla\cdot (e{\bf v})=
-p\nabla\cdot{\bf v} - \nabla\cdot{\bf Q} - n_e^2 \Lambda(T),
\end{eqnarray}
where $\rho$ is the mass density, ${\bf v}$ is the fluid velocity,
${\bf B}$ is the magnetic field vector, ${\bf g}$ is the gravitational
acceleration, $p$ is the gas pressure, $e$ is internal energy density,
${\bf Q}$ is the heat flux, and $\Lambda(T)$ is a cooling function,
$n_e$ is the electron number density, and T is the temperature. We
adopt an equation of state $p = (\gamma -1)e = \rho T /\mu m_p $, with
$\gamma$ = 5/3, adequate for monoatomic gas. We have ignored viscous
terms in the equation of motion in this work; these will be considered
in a future study.

In the absence of magnetic fields, we take the electron thermal
conductivity to be given by its Spitzer value \citep{spitzer62}:
\begin{equation}
\chi=\chi_{\rm s}\approx \frac{1.84\times 10^{-5}T^{5/2}}{\ln \lambda} {\rm erg
}\,\,{\rm s}^{-1}\,{\rm cm}^{-1}\,{\rm K}^{-1}, 
\label{eq5}
\end{equation}
where the Coulomb logarithm $\ln \lambda$ is $\sim 30-40$ for
conditions in a cluster cooling flow.  The local conductive heat flux
will then be given by $ {\bf Q}=-\chi\nabla T$.  If such efficient
conduction operated unimpeded, radiative loses within massive ICM
cores would be more than balanced by the conduction of heat from the
outer portions of the ICM in all but the lowest mass clusters.  This
would eliminate the possibility of a cooling flow or, indeed, any
significant departure from isothermality
\citep{bc81,fabian02,voigt02}.  However, magnetic fields of any
plausible astrophysical strength will strongly suppress transport of
electrons across their line of force.  Thermal conduction is expected
to be efficient along field lines, and suppressed perpendicular to the
field lines.  The form of the heat flux under these conditions is,
\begin{equation}
{\bf Q}=-\chi \,\hat{{\bf b}}\, (\hat{{\bf b}} \cdot \nabla T)
\label{eq_qaniso}
\end{equation}
where $\hat{{\bf b}}$ is a unit vector in the direction of the
magnetic field.  Assuming a spherical temperature gradient, the radial
heat flux can be expressed as  ${\bf Q}\cdot {\bf\hat{r}}= - \chi ({\bf
\hat{b}\cdot\hat{r}})^2 \,\partial T / \partial r $.
For convenience we will define a thermal diffusivity, $\kappa = \chi
T/ p = \kappa_{aniso}\,(n_0/n) (T/T_0)^{5/2}$, where $\kappa_{aniso}$
has dimensions of a diffusion coefficient (${\rm cm^2\,s^{-1}}$), and
$n_0$ and $T_0$ are fiducial values for the electron number density
and temperature, respectively.
\begin{figure*}[t]
\includegraphics[width=0.5\textwidth]{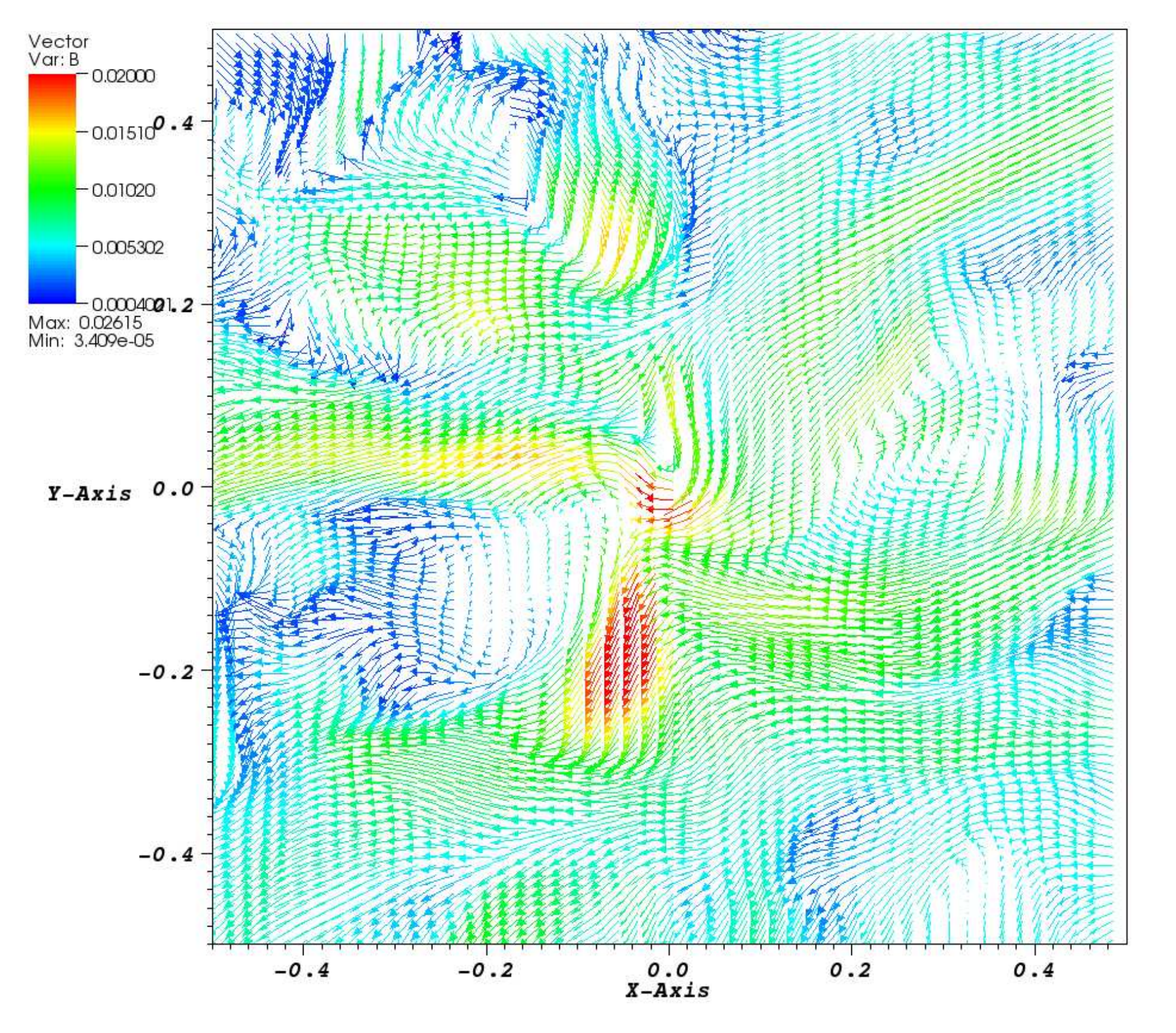} 
\includegraphics[width=0.5\textwidth]{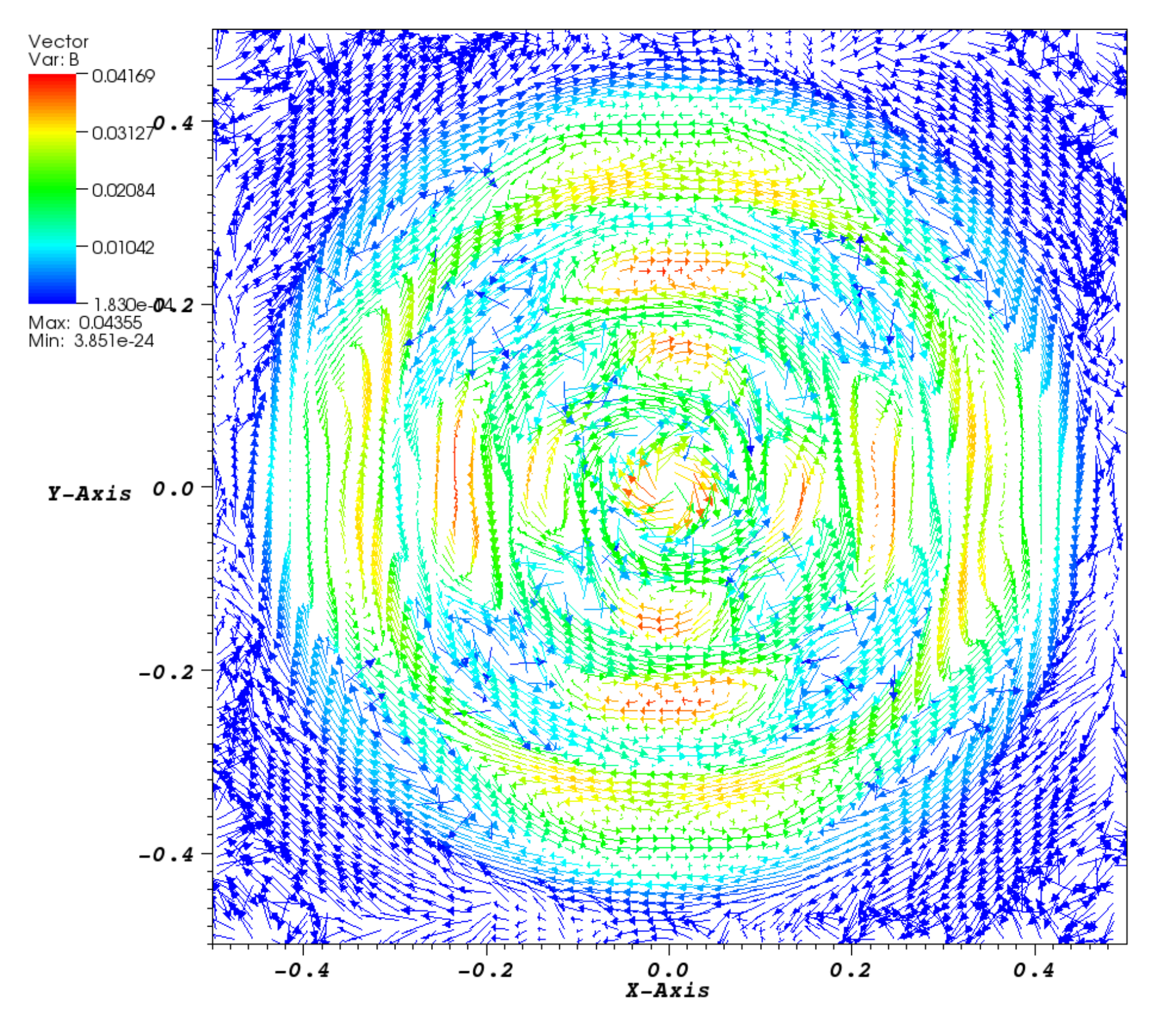}
\caption{Examples of the geometry of magnetic field lines immediately
after the initial radiative relaxation phase in which the cool core is
imprinted into the initial conditions. {\it Left:} Large scale,
loosely tangled magnetic field lines used as a part of initial
conditions in T-models. {\it Right:} Tangled azimuthal field lines
with spatially variable $\phi$ and $\theta$ components used in
A-models. Fields lines in AR-models appear very similar except for the
addition of the split monopole term at the center. Both panels show
the slice through the xy-plane of a cluster center.  Arrows mark the
direction of magnetic field vectors and colors indicate field
strength. The orientation of vectors in the corners of figure
representing A-model is the visualization artifact that arises for
very weak fields.}
\label{initial_fields}
\end{figure*}

Anisotropic thermal conduction is implemented in the {\it Athena} code
via a split operator approach with subcycling \citep{ps05}. Subcycling
ensures stability of the integration and is used whenever the Courant
time step for conduction falls below that used for hydrodynamic
equations. The thermal conduction time step is evaluated as $\delta
t_{cond} = 0.5\, (\delta x)^2 / \kappa_{max}$, where $\kappa_{max} =
\kappa_{aniso}$ for $T_0 = 1$ and $\delta x$ is the size of a
resolution element. The thermal conduction term is implemented using
the method of monotonized central difference limiters \citep{sh07}
which prevents unphysical heat flow from colder to hotter regions
that may arise as a numerical artifact in a variety of algorithms.

In this work, we adopt two forms for the optically-thin radiative
cooling function; both are introduced into the {\it Athena} code using
operator splitting as explicit source terms.  The majority of our
parameter space survey simulations employ a very simple
``bremsstrahlung'' form for the cooling function, $n_e^2
\Lambda=\alpha\, n^2 T^{0.5}$.  We control
the strength of radiative cooling and hence the cooling timescale via
the parameter $\alpha$.  On the other hand, our physical cluster
simulations use the \citet{tn01} approximation for the cooling
function that incorporates the effects of free-free and line cooling:
\begin{equation}
n_e^2\,\Lambda = \left(C_1(k_BT)^{-1.7} + C_2(k_BT)^{0.5}
+C_3\right)n_{i}n_{e},
\label{eq7}
\end{equation}
where $(k_B T)$ is in units of keV and $n_i$ is the ion number
density. For mean metallicity of $Z = 0.3 Z_\odot$, $C_1 =
8.6\times10^{-3}$, $C_2 = 5.8\times10^{-2}$, and $C_3 =
6.4\times10^{-2}$ and the units of $\Lambda$ are $10^{-22} {\rm erg\,
cm^3\, s^{-1}}$. Mean molecular weight corresponding to this
metallicity in case of a near complete ionization is $\mu = 0.59$.

\subsection{Numerical Setup and Initial Conditions\label{S_ics}}

The simulations were performed in a Cartesian coordinate system
$(x,y,z)$ with a cubic spatial domain defined by $x=\pm L/2$, $y=\pm
L/2$, $z=\pm L/2$, where $L=1$. Unless otherwise stated the nominal
numerical resolution used is $100^3$.

All of our simulations have an initial density distribution that is
described by a $\beta$-model, $\rho(r) = \rho_0\,\left(1 + (r/r_0)^2
\right)^{-0.75}$, where $r^2=x^2+y^2+z^2$.  The temperature
distribution is initially isothermal with temperature $T_0$.  We
choose to work in units where $\rho_0=1$, $r_0=0.1$ and $T_0=1$.  The
underlying gravitational potential is assumed to be dominated by the
dark matter, and is static throughout the simulations with the form
\begin{equation}
\Phi = -\frac{3c_s^2}{4\gamma}\ln\left[1+\left(\frac{r}{r_0}\right)^2\right] 
\end{equation}
where $c_s$ is the speed of sound corresponding to temperature $T_0$.
This form corresponds to hydrostatic equilibrium in the initial
isothermal ICM density distribution.

The simplicity of an isothermal hydrostatic atmosphere makes it a
compelling choice of initial condition.  However, the astrophysical
systems of interest have ICM cores that display a positive temperature
gradient.  Furthermore, the interesting dynamics is driven by
temperature gradients.  Thus we employ a pre-cursor simulation where,
starting from the isothermal state, we create a cool core in the
atmosphere by allowing radiative cooling in the absence of conduction.
This continues until the core reaches a temperature of $T_c \approx
0.3 - 0.4T_0$, similar to the range observed in cores.  The density
and magnetic field are allowed to evolve self-consistently during this
initial cooling phase.  We then use this cool-core state as the
starting point for the full simulations (i.e., those including the
anisotropic conduction and associated dynamics).
\begin{deluxetable*}{ccccccc}[t] 
\tablecaption{Value of $\kappa_{\rm aniso}$ in models. \label{table1}}
\tablewidth{0pt}
\tablecolumns{7}
\tablehead{
\colhead{Model} & 
\colhead{{\bf B}-field} & 
\colhead{B1\tablenotemark{\rm a}} & 
\colhead{B2} & 
\colhead{TN\tablenotemark{\rm b}} &
\colhead{B3} &
\colhead{B4} \\
\colhead{} & 
\colhead{structure} & 
\colhead{($\alpha$=0)} &
\colhead{($\alpha$=0.01)} &   
\colhead{($\alpha \approx 0.1$)} &
\colhead{($\alpha$=0.3)} &
\colhead{($\alpha$=1)}  
}
\startdata
  $ 1 $  & T   &        &        &        & 0.1 &    \\
  $ 2 $  & T   & 0.025  & 0.025  &        & 1   & 1  \\
  $ 3 $  & T   & 0.05   & 0.05   &        & 10  & 10 \\
  \hline
   $ 4 $ & A   & 0.01   &        &        &     &    \\
   $ 5 $ & A   & 0.025  & 0.025  & 0.0    & 0.1 &    \\
   $ 6 $ & A   & 0.1    & 0.05   & 0.1    & 1   & 1  \\
   $ 7 $ & A   & 1      &        & 1      & 10  & 10 \\
  \hline
   $ 8 $ & AR &  0.01  & 0.01   &  0.01  &     &    \\
   $ 9 $ & AR & 0.025  & 0.025  &  0.025 & 0.1 &    \\
   $ 10 $ & AR &        &        &  0.1 & 1 & 1\\
   $ 11 $ & AR &        &        &  1     & 10  & 10 \\  
 \enddata
\tablenotetext{a}{B1$-$4: Models with bremsstrahlung cooling function,
$n_e^2\Lambda = \alpha n^2 T^{0.5}$, and $\mu=1$.}
\tablenotetext{b}{TN: Models with Tozzi-Norman cooling function for
metallicity $Z = 0.3\, Z_\odot$ and $\mu=0.59$.}
\end{deluxetable*}

Note that cluster cores initialized in this way are not in
thermodynamic equilibrium and that anisotropic conduction is not
matched to exactly balance radiative cooling at the start of the full
simulation \citep[for alternative approach starting from thermodynamic
equilibrium see][]{pqs09}. While realistic cluster cores are most
likely not in thermodynamic equilibrium either, they experience more
gradual thermal histories compared to that created in our
simulations. We will discuss the effect of this feature later, along
with the description of results.

The initial magnetic field strength is typically chosen so that the
average value of the plasma parameter (i.e. the ratio of thermal
pressure to magnetic pressure) is $\beta = 8\pi p/ B^2 \sim 10^2-
10^3$.  We explore several different
scenarios for the initial structure of magnetic fields. Firstly, we define field line geometry 
given by the form,
\begin{eqnarray}
B_x  &=&  \frac{B_0}{\sqrt 3} \cos(\pi\,y) \cos(\pi\,z),\\
B_y  &=&  \frac{B_0}{\sqrt 3} \cos(\pi\,x) \cos(\pi\,z),\\
B_z  &=&  \frac{B_0}{\sqrt 3} \cos(\pi\,x) \cos(\pi\,y),
\label{eq9_11}
\end{eqnarray} 
This initial geometry is self-consistently evolved within an isothermal core 
until field lines assume loosely tangled geometry while retaining the
large scale character given by the initial form. This field configuration is then 
used as a part of the initial conditions. These models are representative of an 
idealized scenario in which anisotropic conduction initially operates efficiently by
transporting heat from the outer, hotter regions of a cluster to the
cooler core center. Hereafter, we denote this class of models by ``T''. In this model, the
constant $B_0$ is the initial amplitude of the magnetic
field. Secondly, we examine models with tangled purely azimuthal field
lines, i.e., a field structure that is initially wrapped onto surfaces
of constant $r$.  In this scenario we
investigate the evolution of cool cores when the initial field
geometry is set up to inhibit heat conduction towards the core. We use
these models to test the hypothesis that spherical collapse and MHD
turbulence driven by the heat flux instability can regulate field-line
insulation and drive a reverse convective thermal flux
\citep{br08}. Defining a spherical polar coordinate system
$(r,\theta,\phi)$ where $\theta=0$ is aligned along the $z$-axis, we
employ an azimuthal field structure defined by,
\begin{eqnarray}
B_{\theta} & = & 2B_0 (1 + \sin(2\pi r/r_1)) \sin\theta \cos(2\phi),\\
B_{\phi} & = & 2B_0 (1 + \sin(2\pi r/r_2)) \sin(3\theta) - \nonumber \\ 
  &  & B_0 (1 + \sin(2\pi r/r_1)) \sin(2\phi) \sin(2\theta),\\
B_r&=&0.  
\label{eq12_14}
\end{eqnarray} 
We refer to these as ``A'' models.  Here, $B_0^2 = 8\pi\,p/\beta$,  
$r_1 = 0.1$ and $r_2 = 0.087$ are coherence lengths defining 
characteristic scales on which magnetic
field vector changes direction.  Finally, we examine a modification of
the tangled azimuthal field model in which we include a radial
component in the form of a split monopole,
\begin{eqnarray}
B_r & = & sign\, \frac{B_{r0}}{r^2 + \epsilon}, \;\;  sign =\left\{ \begin{array}{ll}
  1  & \mbox{if $z > 0$} \\
 -1  & \mbox{otherwise}
  \end{array} \right. \label{eq15}
\end{eqnarray}
We shall refer to these as our ``AR'' models. The constant $\epsilon
<< r_0$ is a small number chosen to avoid a singularity at the origin.
We use this simple setup to study cases in which magnetic field at the very center of a
cool core may be tangled and enhanced in the process of spherical
collapse. We test to which extent such fields provide a local magnetic 
support as well as channel the heat within a small volume of the core.
All initial magnetic field geometries are self-consistently evolved
during the initial precursor simulation in which the cool core is
imprinted into the initial condition.  For illustration, slices
through the ``T'' and ``A'' field geometries (at a time just after the
initial radiative relaxation of the model cluster) are shown in
Fig.~\ref{initial_fields}.

\begin{figure}[t]
\includegraphics[width=0.5\textwidth]{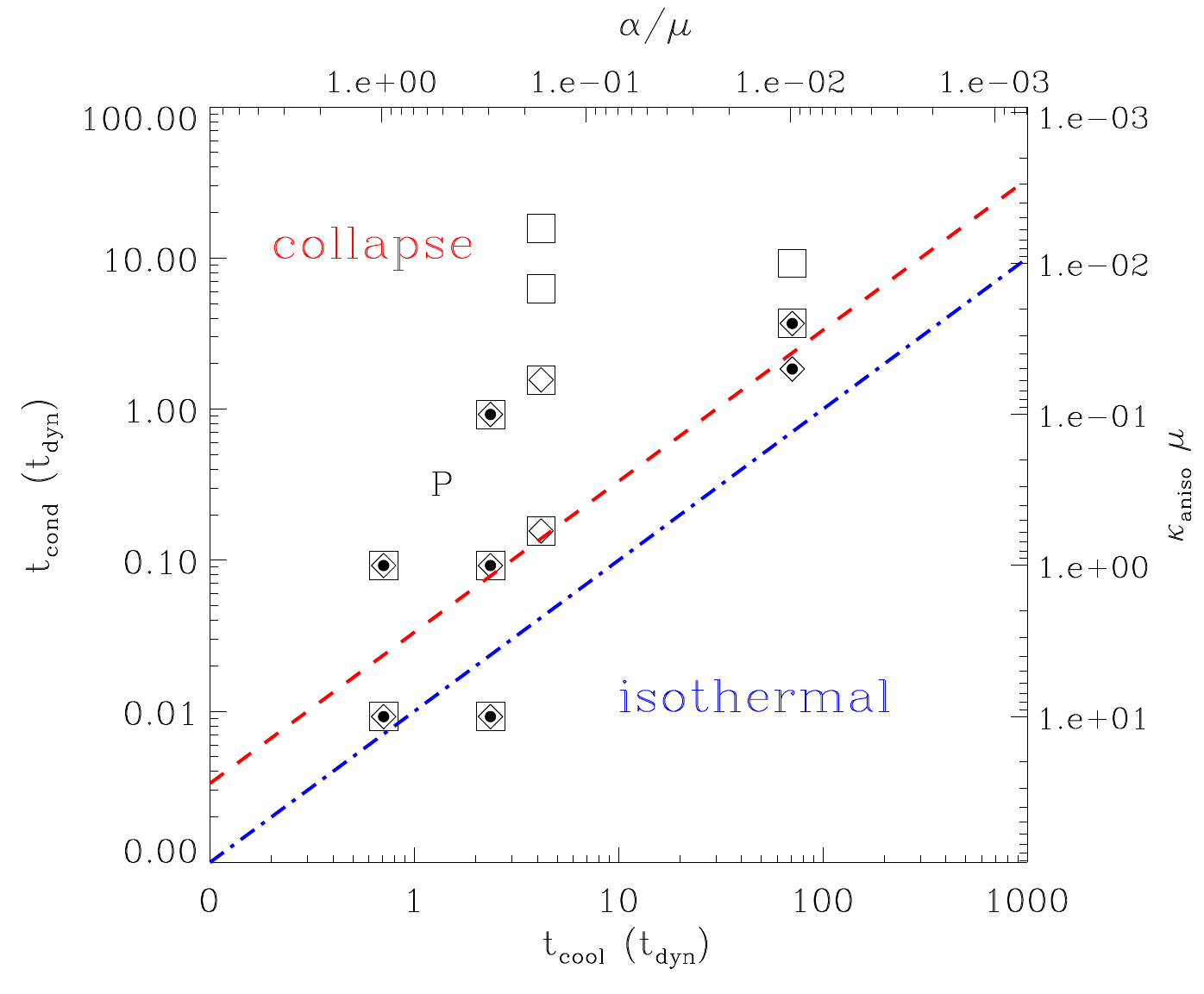}
\caption{Illustration of the parameter space of clusters explored in
this study shown in terms of cooling and heat conduction time,
normalized to the dynamical time. Also marked are corresponding values
of $\kappa_{\rm aniso}$ and $\alpha$ (for a list of values used in
simulations see Table~\ref{table1}). Filled circles, diamonds, and
squares represent T, A, and AR-models, respectively. A combination of
symbols signifies that different magnetic field configurations were
simulated at the same value of $\alpha$ and $\kappa_{\rm
aniso}$. Letter ``P'' marks the location of the Perseus-like
model. The investigated parameter space includes but it is not limited
to that of realistic clusters which occupy the range $1 \lesssim
t_{\rm cool}/ t_{\rm dyn} \lesssim 10$ and ${\rm few}\times 0.1
\lesssim t_{\rm cond}/ t_{\rm dyn} \lesssim 10$.
\label{param_space}} 
\end{figure}

We complete the specification of our simulations by describing the
boundary conditions.  The velocity boundary conditions at the
interface of the active and ghost zones are zero-gradient outflow, and
the pressure and density of the gas are extrapolated from the active
zone into the ghost zone so as to maintain hydrostatic equilibrium in
the ghost zones.  This choice of boundary conditions ensures stability
and prevents spurious sound waves from developing at the boundaries.
The temperature is fixed to the virial temperature $T_0$ on a
spherical shell at a radius $r=L/2$.  Thus, while the dynamics are
followed in the full cubical domain, only the interior of the $r=L/2$
sphere is physically interesting.  Early experimentation with applying
temperature boundary conditions at the edge of the cubic domain
revealed behavior that appeared to depend upon the overall orientation
of our Cartesian domain.  Thus, we chose to impose thermal boundary
conditions on a spherical surface in order to emulate hot intercluster
plasma surrounding the cool core.

\subsection{Characteristic Time Scales\label{S_timescales}}

For our parameter space survey, we catalogue our simulated clusters
according to their characteristic dynamical, conduction, and cooling
timescales.  These are defined (in code units) by,
\begin{eqnarray}
t_{\rm dyn} & = & \frac{R}{c_s} \approx \frac{0.39}{\sqrt{\mu}} \,\, , \label{eq16} \\ 
t_{\rm cond} & = & \frac{R^2}{\kappa}\frac{T_0}{\Delta T} \approx
\frac{0.036}{\kappa_{\rm aniso}\,\mu} ,\label{eq17} \\
 t_{\rm cool} & = & \frac{e}{n_e^2\Lambda} \approx 0.27\,\frac{\mu}{\alpha}\,\, \label{eq18}
\end{eqnarray}
where $t_{\rm dyn}$ is calculated for the cooling core radius $R = L/2
= 0.5$ and the average speed of sound in the region with virial
temperature $T_0 = 1$, $\gamma = 5/3$, and $\mu =1$ or 0.59, as
specified for a given run. When calculating $t_{\rm cool}$ we
evaluated the shortest time scale corresponding to the core center
where typically $T_c \approx 0.3$ and $\rho_c \approx 3$ at the
beginning of a simulation. In eq.~(\ref{eq17}), $\kappa$ was evaluated
for an average density in the core region $\langle\rho\rangle=0.1$ and
virial temperature.\footnote{Note that the expression commonly used in
literature to estimate $t_{\rm cond}$ omits factor $T_0/\Delta T$ and
thus, refers to the shortest timescale for heat conduction.} According
to the theory of linear growth applied to the weak magnetic field
regime, the plasma is unstable to HBI on the local dynamical time
whenever the vector of magnetic field is aligned with the temperature
gradient and unbridled conduction is allowed in the direction of
gravity. In the limit when lines of magnetic field are preferentially
oriented across the temperature gradient, this condition is modified
by a factor $({\bf \hat{b}\cdot\hat{r}})$ in such way that
\begin{equation}
t_{\rm HBI} =  \left( \frac{d \ln T}{dr} \frac{d\Phi}{dr}\right)^{-1/2} \left|{\bf \hat{b}\cdot\hat{r}}\right|^{-1}  \,\, ,
\label{eq_HBItime}
\end{equation}
\citep{quataert08}. For the values of initial magnetic field used in
simulations the Alfv{\'e}n time scale is typically much longer with
respect to the above time scales and thus, Alfv{\'e}n waves are not
expected to play a significant role during the phase characterized by
the HBI instability and radiative cooling.

\section{Results\label{S_results}}

We have performed an extensive suite of simulations.  Our primary set
of 36 simulations (our parameter space survey simulations) represent a
systematic exploration of the dependence of the cluster core evolution
on ratios of the characteristic timescales (i.e., $t_{\rm cool}/t_{\rm
dyn}$ and $t_{\rm cond}/t_{\rm dyn}$), and the magnetic field
geometry.  The cooling and conduction timescales are controlled via
the parameters $\alpha$ and $\kappa_{\rm aniso}$.  Table~1 defines this
set of models.  We explore three field geometries (the T, A, and AR
geometries of Section~\ref{S_ics}) and five cooling laws (B1--4$+$TN;
this includes the zero-cooling case $\alpha=0$).  For each choice of
field geometry and cooling law, 2--4 simulations are performed with
different degrees of thermal conduction.  Thus, we can map out the
behavior of clusters as a function of their position on the 2-d
parameter space $(t_{\rm cool}/t_{\rm dyn}, t_{\rm cond}/t_{\rm dyn})$
and, furthermore, assess whether the initial field geometry can
influence/change the eventual state of the cluster for a given
position on the $(t_{\rm cool}/t_{\rm dyn}, t_{\rm cond}/t_{\rm
dyn})$-plane.  We refer to individual models from Table~\ref{table1}
according to their alphanumeric tag consisting of the letter
abbreviation corresponding to the magnetic field structure (T, A, or
AR), a model number (1$-$11), and the descriptor of the cooling
function (B1$-$B4 or TN). According to this convention a model with
purely azimuthal initial magnetic field geometry, $\kappa_{\rm aniso}
= 0.025$, and no radiative cooling is marked as A5B1. {\it Note that this
study includes but it is not limited to the part of the parameter
space occupied by realistic clusters.}

\begin{deluxetable*}{ccrrrrr}[t] 

\tablecaption{Time scale (in $t_{\rm dyn}$) for core evolution towards
collapse (C) or isothermal state (I). \label{table2}}

\tablewidth{0pt}
\tablecolumns{7}
\tablehead{
\colhead{Model} & 
\colhead{{\bf B}-field} & 
\colhead{B1} & 
\colhead{B2} & 
\colhead{TN} &
\colhead{B3} &
\colhead{B4} \\
\colhead{} & 
\colhead{structure} & 
\colhead{($\alpha$=0)} &
\colhead{($\alpha$=0.01)} &   
\colhead{($\alpha \approx 0.1$)} &
\colhead{($\alpha$=0.3)} &
\colhead{($\alpha$=1)}  
}
\startdata
   $ 1 $  & T  &             &             &             & 12 (C)    &        \\
   $ 2 $  & T  & 70 (I)      & $>1000$ (C) &             & 80 (C)    & 10 (C)   \\
   $ 3 $  & T  & 60 (I)      & 80 (I)      &             &  $<3$ (I) & $<3$ (I) \\
  \hline
   $ 4 $  & A  & $> 500$ (I) &             &             &           &          \\
   $ 5 $  & A  & 500 (I)     & 380 (C)     & 30  (C)     & 8  (C)    &          \\
   $ 6 $  & A  & 240 (I)     & 510 (C)     & 160 (C)     & 18 (C)    &  5 (C)   \\
   $ 7 $  & A  & 50  (I)     &             & $> 700$ (I)\tablenotemark{\rm a} & 20 (I) & 13 (C)   \\
  \hline 
   $ 8 $  & AR & $> 500$ (I) & 420 (C)     &  90  (C)    &           &          \\
   $ 9 $  & AR & 500 (I)     & 560 (C)     &  110 (C)    & 8 (C)     &          \\
   $ 10 $ & AR &             &             &  230 (C)    & 20 (C) & 5 (C) \\
   $ 11 $ & AR &             &             & $> 700$ (U)\tablenotemark{\rm b} & 20 (I) & 25 (C)\\  
 \enddata

\tablenotetext{a}{A7TN: Evolution of the core that initially headed
towards collapse was reversed to evolution towards isothermal state.}

\tablenotetext{b}{AR11TN: Final state for this model appears 
undetermined (U) at the end of the simulation.}

\end{deluxetable*}

We expect the HBI to be active; hence, one could readily envisage a
scenario in which field line reorientation (by the HBI) completely
insulates the central region from the conductive heat flux, which
would prevent complete equilibration of core.  However, as we shall
see, while field line reorientation does occur, it never completely
insulates the core. Figure~\ref{param_space} shows the models on the
2-d parameter space $(t_{\rm cool}/t_{\rm dyn}, t_{\rm cond}/t_{\rm
dyn})$.  The results of our simulations clearly divide this parameter
space into three regions depending upon the ratio $t_{\rm cool}/t_{\rm
cond}$.  If $t_{\rm cool}/t_{\rm cond}\gtrsim10^2$, cooling appears to
be too weak to sustain any significant temperature gradient against
the action of conduction, and all simulated cluster cores become
approximately isothermal.  On the other hand, if $t_{\rm cool}/t_{\rm
cond}\lesssim25$, our simulated clusters always undergo a cooling
catastrophe although it can be appreciably delayed by the action of
HBI-regulated conduction.  In the intermediate regime ($25\lesssim
t_{\rm cool}/t_{\rm cond}\lesssim 10^2$), clusters can evolve to
either an isothermal state or undergo a cooling catastrophe depending
upon the initial magnetic field configuration.  These are the only
outcomes; none of our simulated clusters achieved a non-isothermal
quasi-steady state.

In the rest of this section we shall describe these results in detail
and, in particular, examine the role of the HBI in influencing the
thermal evolution of the cluster core.

\subsection{Clusters with an isothermal final state\label{S_isothermal}}

As an extreme case, we first describe the evolution of cool cores in
models without radiative cooling, $\alpha = 0$ (also referred to as
``B1'' models).  The fact that these clusters evolve to an isothermal
state is not as trivial as it first appears.  Recall that our full-up
simulations start with a cool core (and hence a positive temperature
gradient) already imprinted on the ICM atmosphere.  We expect the HBI
to be active and, hence, one could readily envisage a situation in
which field-line re-orientation by the HBI completely insulates the
core from the conductive heat flux and hence prevents complete
equilibration of the core.  As we will see, field-line re-orientation
does indeed occur but never completely insulates the core.

\begin{figure*}[t]
\center{
\includegraphics[width=0.45\textwidth]{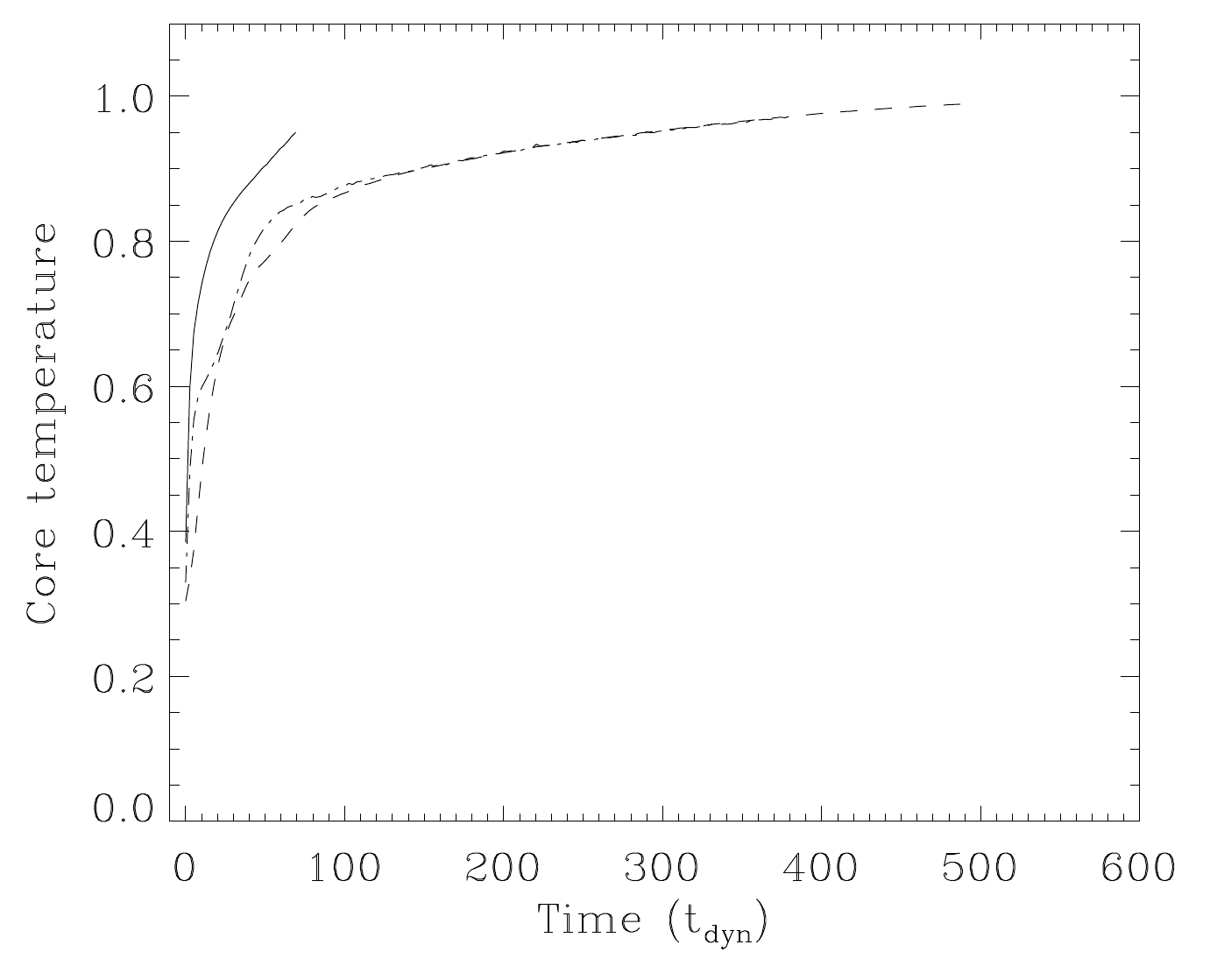} 
\includegraphics[width=0.45\textwidth]{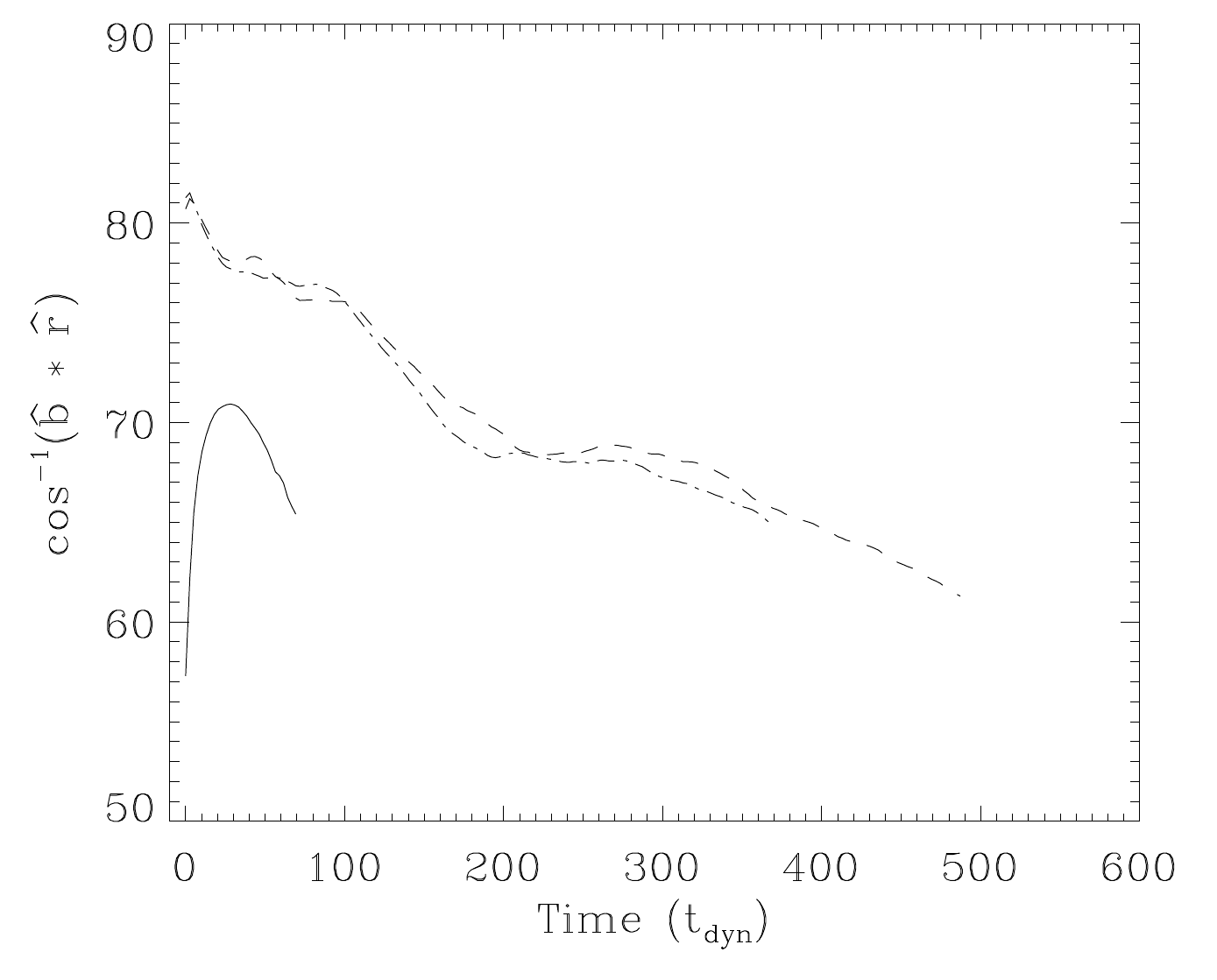}}
\caption{{\it Left:} Evolution of the core temperature in T2B1
(solid), A5B1 (dashed), and AR9B1 (dash-dot) models,
respectively. {\it Right:} Evolution of the mean angle,
$\langle\theta_B\rangle$, between the lines of magnetic field and the
radius unit vector (same models), indicating the alignment of the
field lines with the global temperature gradient. \label{fig3}}
\end{figure*}

Table~\ref{table2} (B1 column) shows the time taken for each of the
non-cooling models to achieve an approximately isothermal state.
Cores with loosely tangled field lines, captured in T-models, evolve
to isothermal state on significantly shorter time scales with respect
to the cores in models A and AR, where the field line geometry is
preferentially azimuthal.  Figure~\ref{fig3} shows the evolution of
the core temperature and mean angle between magnetic field lines and
the radius unit vector, $\hat{r}$, in three models representing
different magnetic field geometries, T2B1, A5B1, and AR9B1, all at the
same value of $\kappa_{\rm aniso} = 0.025$. The core temperature was
measured at the center of the core, whereas the angle $\langle
\theta_{\rm B}\rangle$ is calculated as a volume average of $
\cos^{-1}(\hat{b}\cdot\hat{r})$, and it indicates the alignment of the
field lines with the global temperature gradient\footnote{The region
of the cluster core outside of $0.3\,L$ is not taken into account in
the calculation of $\langle\theta_{\rm B}\rangle$, in order to focus
on the field structure closer to the core center and avoid effects of
the boundaries.}. As expected, the action of the HBI is most apparent
in the tangled field model (T2B1) where we see a rapid reorientation
of the field lines from $\langle\theta_{\rm B}\rangle \approx
57^\circ$ to $\langle\theta_{\rm B}\rangle \approx 71^\circ$.  The
corresponding effective thermal conductivity is $({\bf
\hat{b}\cdot\hat{r}})^2\approx 0.25$ of the Spitzer value but
decreases as the field lines are re-oriented.  Once the core has
equilibrated, the HBI is no longer driven and the field lines become
more disoriented again leading to a final value between
$\langle\theta_{\rm B}\rangle \approx 60^\circ - 70^\circ$.  A similar
final field orientation results in the A- and AR-models (i.e., the
azimuthal initial field configurations).  In these cases, the core
takes substantially longer to equilibrate, however, with the initial
heat flux corresponding to only $\sim0.02$ of the Spitzer value.  In
comparison, the time scale for unbridled conduction in this set of
models is $t_{\rm cond}\approx 3.7\,t_{\rm dyn}$.

\begin{figure*}[t]
\center{
\includegraphics[width=0.45\textwidth]{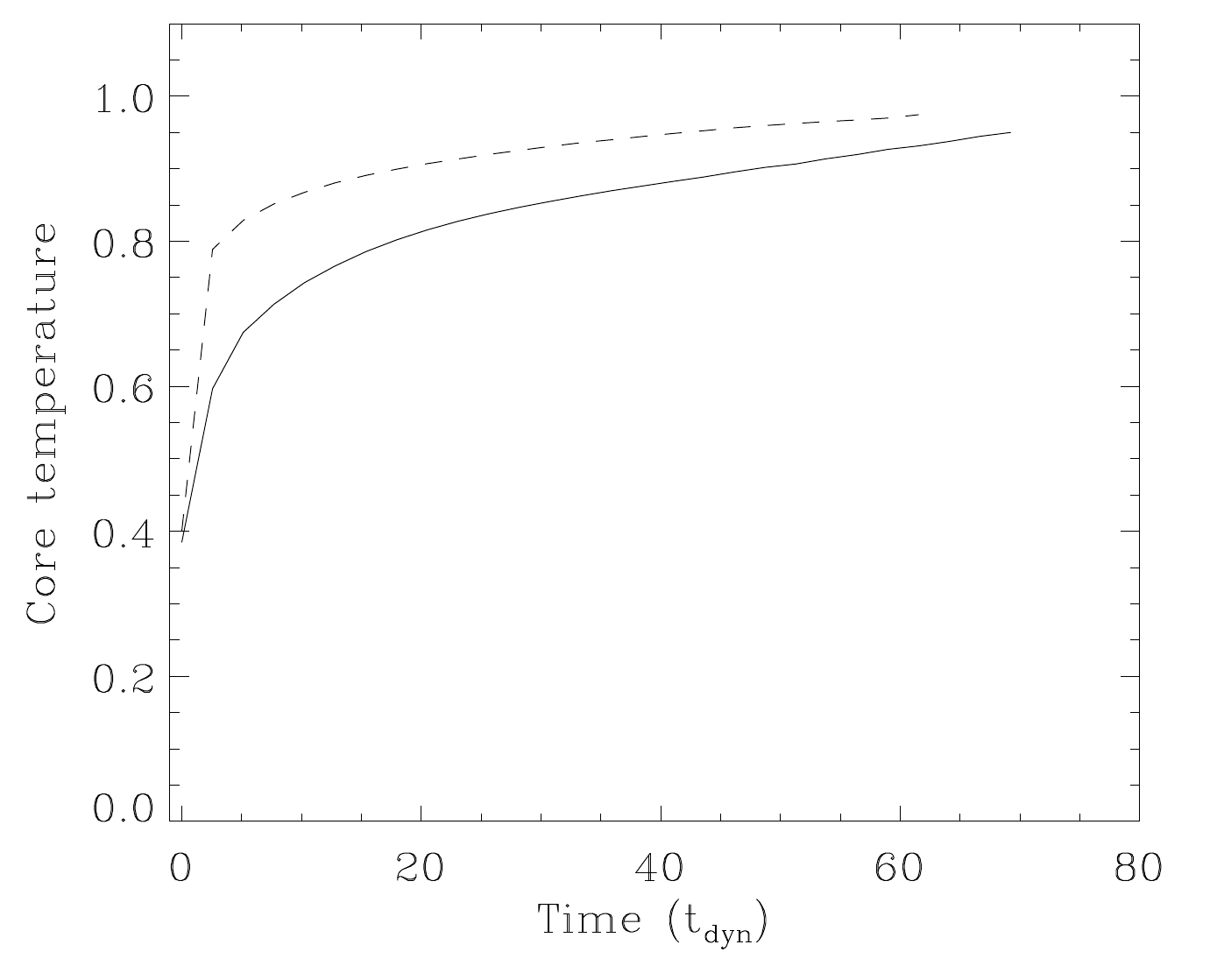} 
\includegraphics[width=0.45\textwidth]{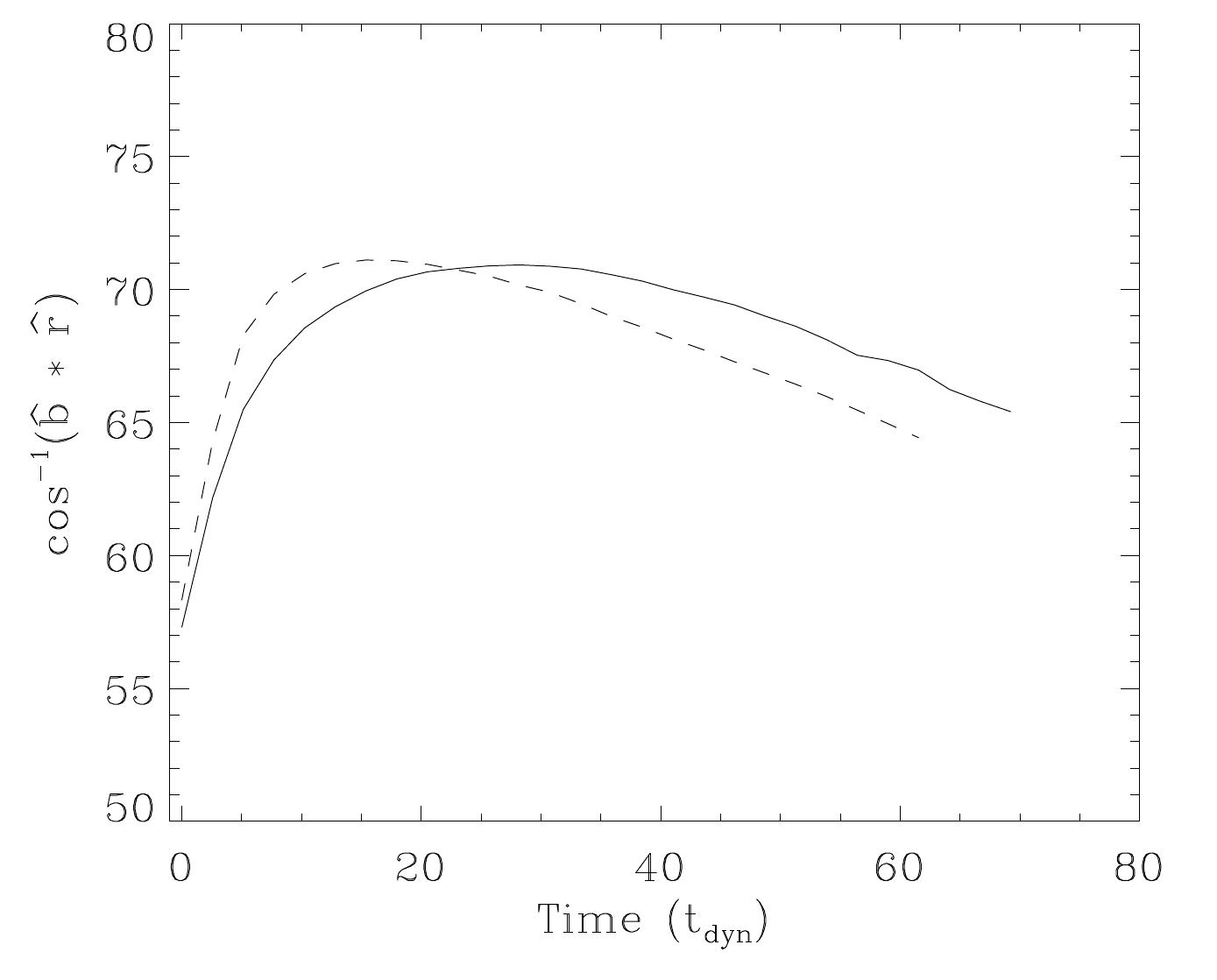}}
\caption{{\it Left:} Evolution of the core temperature in T2B1 (solid)
and T3B1 (dashed) models, respectively. {\it Right:} Evolution of the
$\langle\theta_B\rangle$  for the same two models. \label{fig4}}
\end{figure*}

\begin{figure*}[t]
\center{
\includegraphics[width=0.45\textwidth]{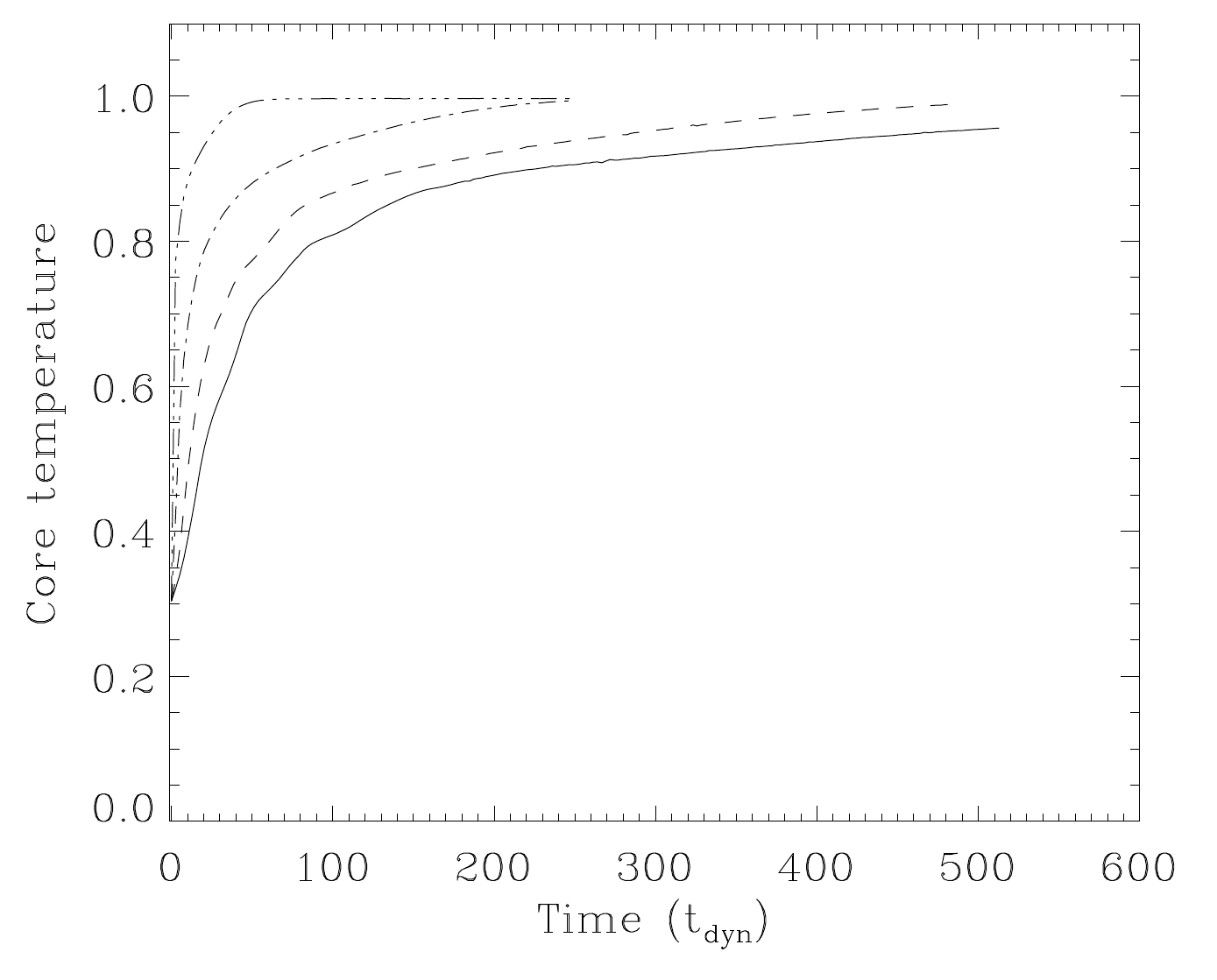} 
\includegraphics[width=0.45\textwidth]{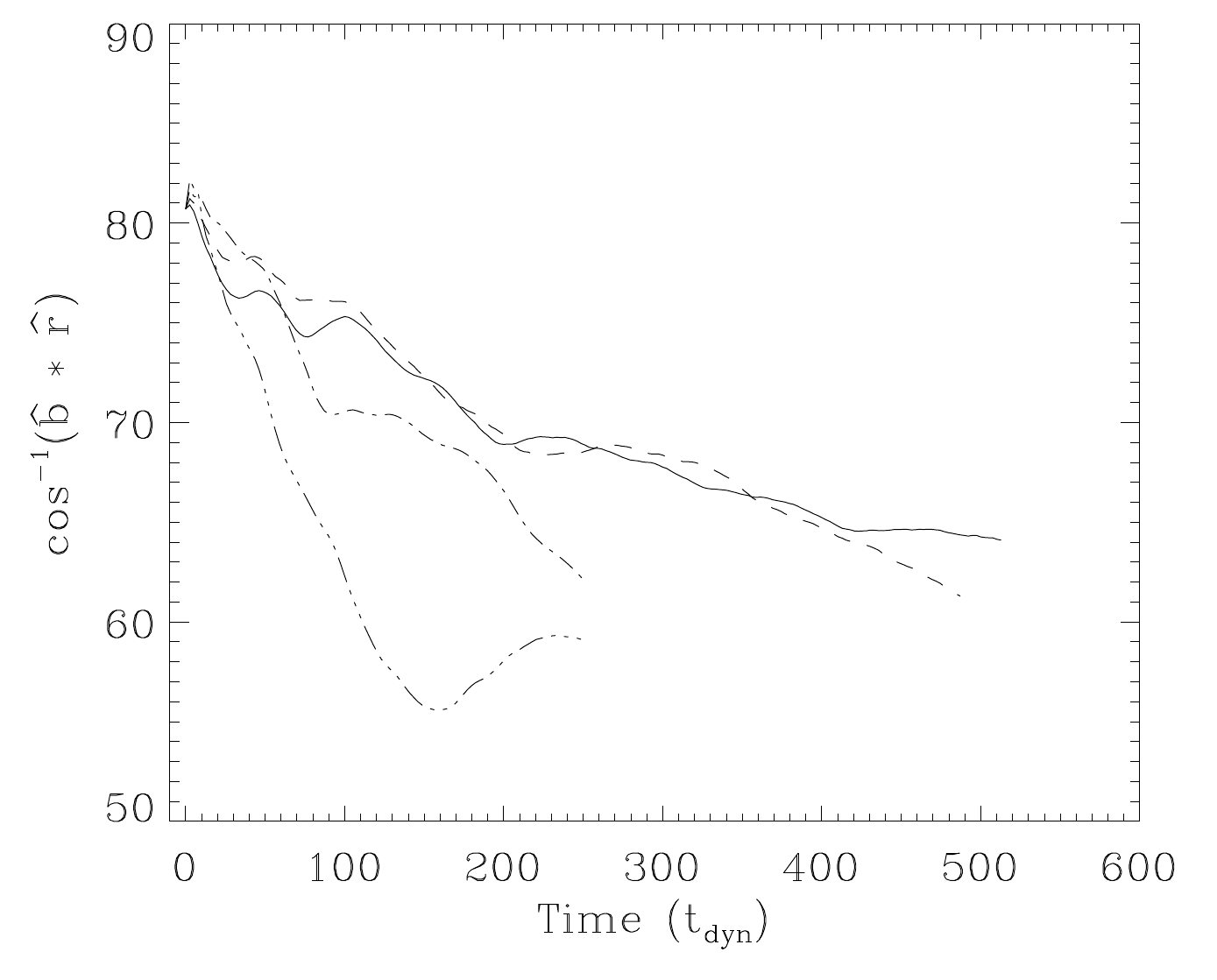}}
\caption{{\it Left:} Evolution of the core temperature in four
A-models: A4B1 (solid), A5B1 (dashed), A6B1(dash-dot) and A7B1
(dash-3-dot), respectively. {\it Right:} Same four models, evolution
of $\langle\theta_B\rangle$. \label{fig5}}
\end{figure*}

Given the same initial geometry of magnetic field lines, the cores
with higher values of $\kappa_{\rm aniso}$ evolve towards
isothermality on shorter time scales. The behavior of models shown in
Figure~\ref{fig4} and Figure~\ref{fig5} confirms this general
expectation, summarized in eq.~(\ref{eq17}). Figure~\ref{fig4} shows
the evolution of the core temperature and orientation of the magnetic
field lines in T-models, T2B1 and T3B1. Similarly, Figure~\ref{fig5}
shows the evolution for four A-models, A4B1, A5B1, A6B1, and A7B1.

\begin{figure*}[t]
\center{
\includegraphics[width=0.48\textwidth]{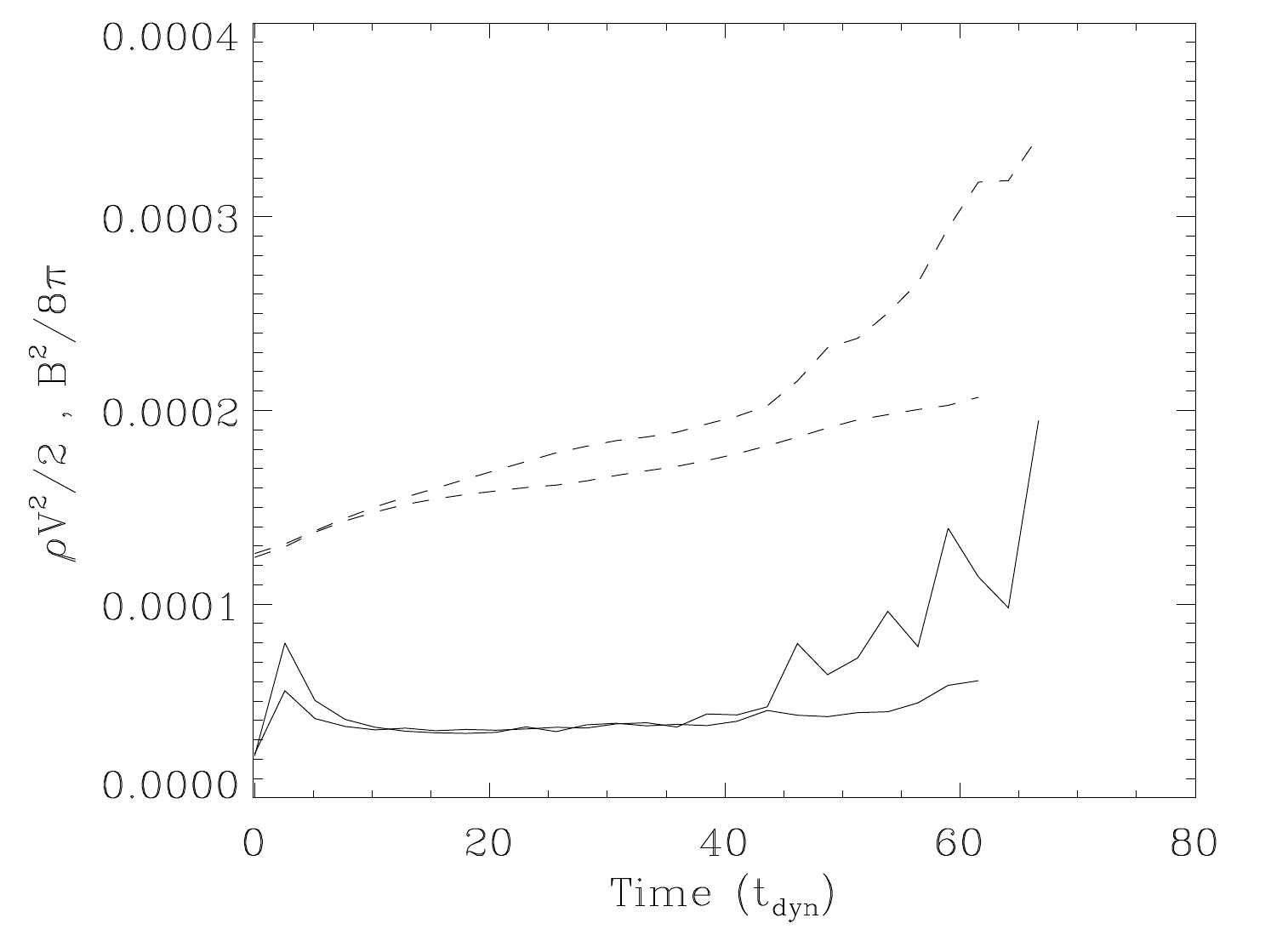} 
\includegraphics[width=0.49\textwidth]{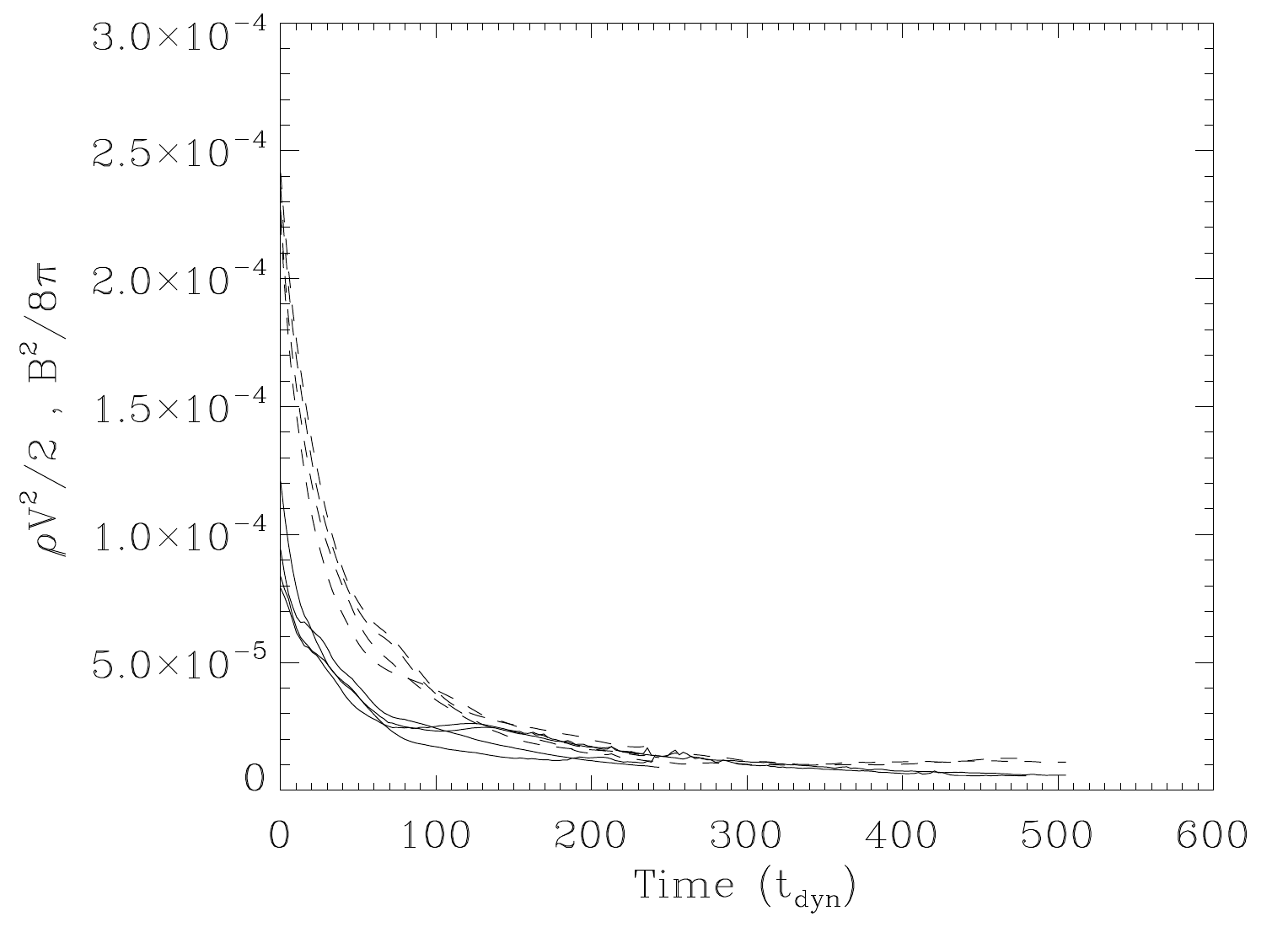}}
\caption{Evolution of the kinetic (solid) and magnetic (dashed) energy
density in two T-models shown in Figure~\ref{fig4} (left) and four
A-models from Figure~\ref{fig5} (right) without radiative
cooling. Both components of energy density are normalized to the
instantaneous internal energy density. \label{fig6}}
\end{figure*}

Interestingly, these non-cooling cores remain rather kinematically
quiescent throughout the entire evolution, and magnetic field does not
play a significant role in subsequent dynamical evolution of the core.
Figure~\ref{fig6} shows that the early evolution of the magnetic and
kinetic energy density exhibit different behaviors in T- and A-models.
In the T-models, the volume averaged magnetic and kinetic energy
increase over time due to the action of the HBI; it appears that a
weak and short lived dynamo may be established.  The A- and AR-models,
on the other hand, show a monotonic decrease in the kinetic and
magnetic energy densities as the non-equilibrium azimuthal magnetic
fields relax and numerically reconnect.  In both panels the components
of energy density are normalized to the internal energy density of the
gas and, hence, the kinetic and magnetic energy represent a small
fraction ($\sim 10^{-4}$) of the internal energy. This implies that
cores are kinematically quiescent and that even after it has been
enhanced (T-models), magnetic field is not sufficiently strong to be
dynamically important.


In fact, these qualitative results are borne out even when there is
cooling present provided it is weak and much slower than the heat
conduction in the sense that $t_{\rm cool}/t_{\rm dyn}>10^2$.
Examples of such behavior are models T3B3, A7B3, and AR11B3, all of
which evolve towards isothermal final state.

\subsection{Clusters which undergo catastrophic collapse\label{S_collapse}}

Real cooling core galaxy clusters do not have isothermal ICM
atmospheres and hence, while this avoids the cooling catastrophe, it
must not be considered a physical outcome.  The more interesting case
are those clusters in which cooling is too effective to allow an
isothermal state to be achieved.  As already mentioned, all of our
model clusters with $t_{\rm cool}/t_{\rm dyn}<25$ undergo an eventual
cooling catastrophe; Table~\ref{table2} lists the time taken for the
model clusters to either undergo catastrophic collapse (C) or
equilibrate to an isothermal state (I).  We now discuss the collapsing
clusters in more detail.

\begin{figure*}[t]
\center{
\includegraphics[width=0.45\textwidth]{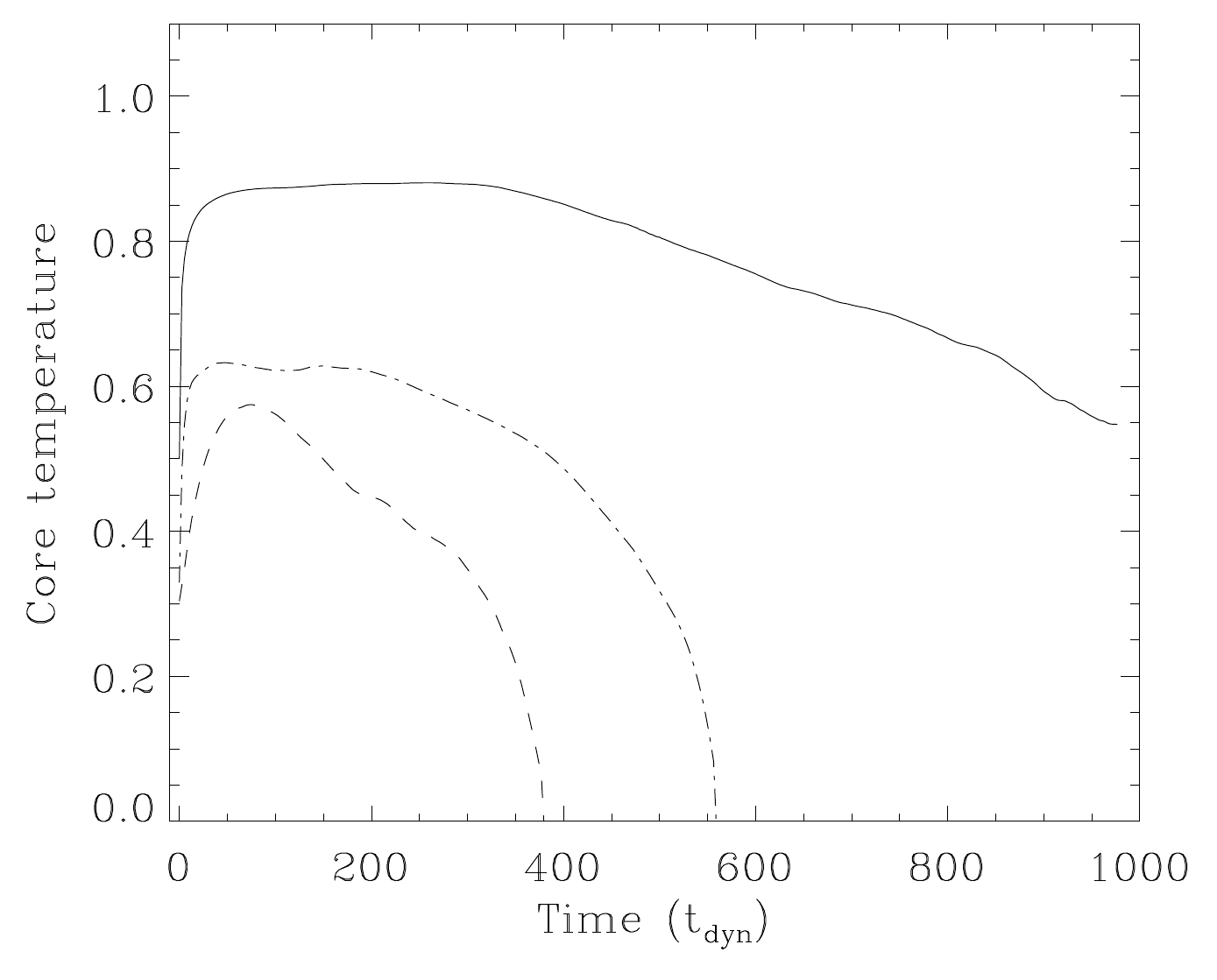} 
\includegraphics[width=0.45\textwidth]{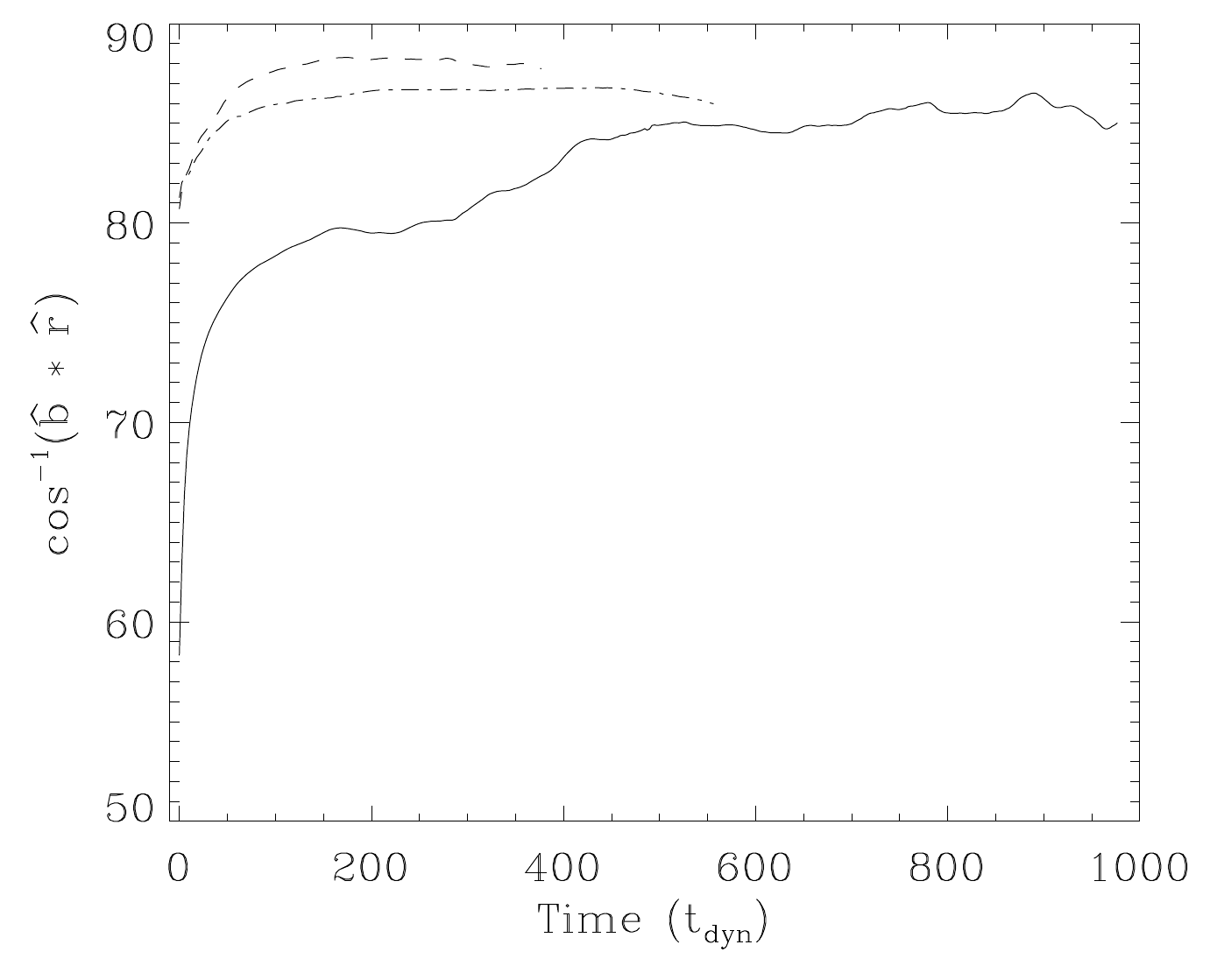}}
\caption{{\it Left:} Evolution of core temperature in models with
different magnetic field structure, all with $\kappa_{\rm
aniso}=0.025$ and $\alpha=0.01$: T2B2 (solid), A5B2 (dashed), and
AR9B2 (dash-dot). The nominal cooling time scale for this group of
models set by value of $\alpha$ is $t_{\rm cool}\approx 69\,t_{\rm
dyn}$. {\it Right:} Same models, evolution of magnetic field
orientation.
\label{fig7}} 
\end{figure*}

\begin{figure}[t]
\includegraphics[width=0.5\textwidth]{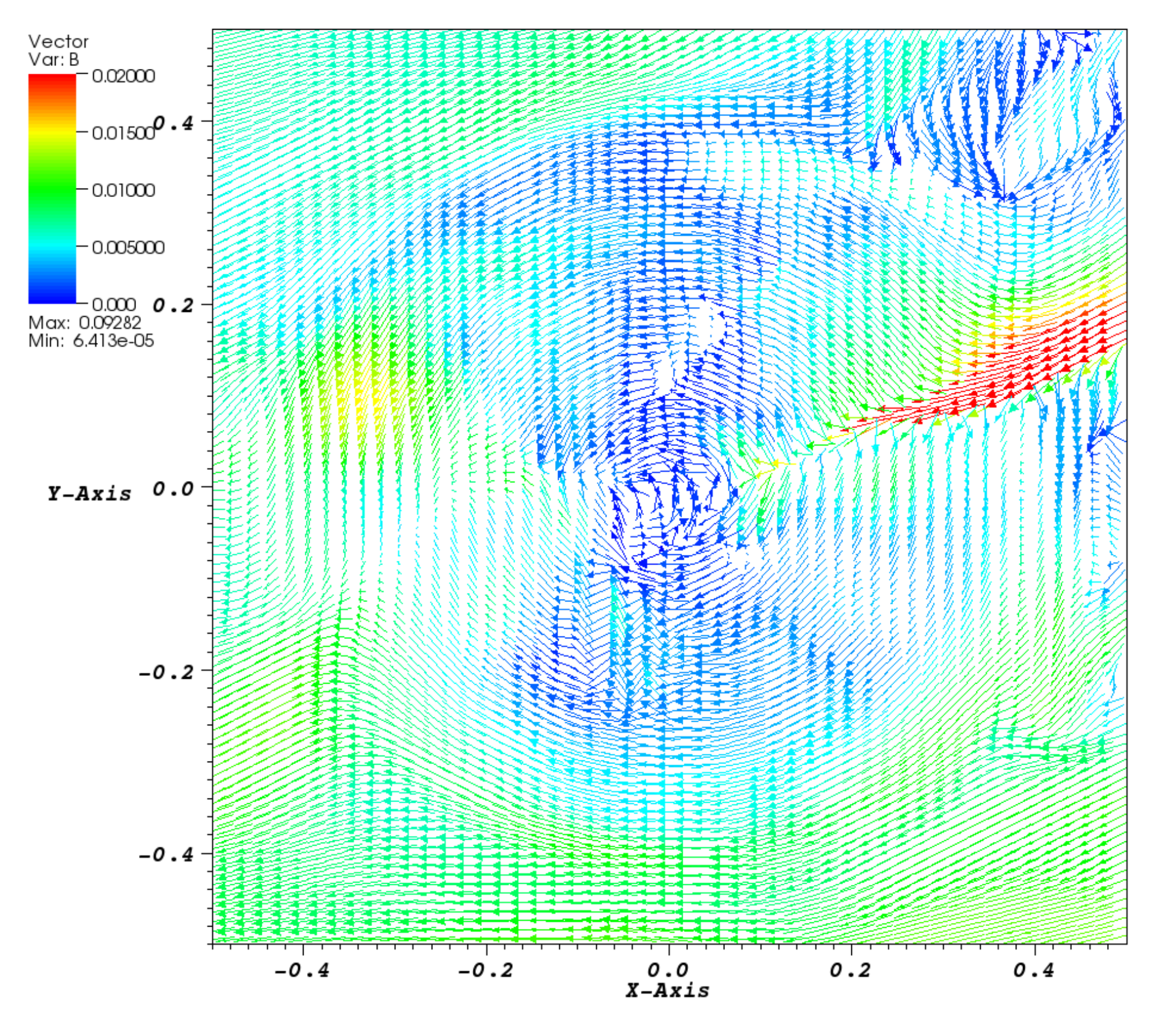}
\caption{Final orientation of the magnetic field lines in model T2B2,
800 dynamical times after the beginning of the simulation. Large
scale, loosely tangled magnetic field lines with $\langle\theta_{\rm
B}\rangle = 58^\circ$ are used as a part of the initial conditions in
this model, as illustrated in the left panel of
Figure~\ref{initial_fields}.  Traces of the initial topology are still
present at late times in the strongest component of the field,
however, most of the lines are wrapped around the core as a
consequence of HBI. Final value of $\langle\theta_{\rm B}\rangle =
85^\circ$. Figure shows a slice through the xy-plane of a cluster
center.
\label{t2b2_field}}
\end{figure}

To illustrate many of the dynamical aspects of these collapsing
clusters, we discuss run T2B2 in detail.  Figure~\ref{fig7} (solid
line) shows the evolution of the core temperature together with the
average field line orientation for run T2B2.  At $t=0$, the core is
not in a state of thermal balance, with conductive heating
overwhelming the radiative cooling.  The core is rapidly heated from
$T\approx 0.4T_0$ almost to a state of isothermality ($T\approx
0.85T_0$).  The HBI is clearly playing a role during this early
heating event, as revealed by the rapid re-orientation of the field
geometry towards an azimuthal configuration (Fig.~\ref{t2b2_field}).
After this early heating event, an approximate balance between
conductive heating and radiative cooling is maintained for a duration
of $\sim 300t_{\rm dyn}$.  Eventually, azimuthal field line wrapping
reduces the conductive heat flux to the point where the core
temperature starts to decrease.  This slow collapse will end in a
cooling catastrophe, in this case at some time $t>1000t_{\rm dyn}$.
For comparison, the nominal cooling time scale for this group of
models set by the value of $\alpha$ and calculated according to the
equation~\ref{eq18} is $t_{\rm cool}\approx 69\,t_{\rm dyn}$.  Thus,
we see that thermal conduction can dramatically increase the time
taken for the cluster to undergo the cooling catastrophe even when the
HBI is inducing azimuthal field-line wrapping within the cluster core.

As may be expected, cores with initially loosely tangled fields
(T-models) evolve towards collapse on time scales longer than cores
with initially azimuthal fields (A- and AR-models)\footnote{Even in
the A-model, the field configuration at the start of the full
simulation is not perfectly azimuthal due to field distortions
imprinted during the precursor simulation in which the cool core is
formed.}.  Furthermore, the AR-model clusters (which include a split
monopole term to the initial field configuration and hence possess a
significant radial field in the core regions) have a collapse that is
delayed with respect to A-models.  Figure~\ref{fig7} compares the
evolution of core temperature and $\langle\theta_{\rm B}\rangle$ in
models T2B2, A5B2, and AR9B2, characterized by different initial
magnetic field structure and identical cooling and conduction
($\kappa_{\rm aniso}=0.025$ and $\alpha=0.01$).  In A and AR-models,
the field orientation starts at $\langle\theta_{\rm B}\rangle\approx
80^\circ$; this permits sufficient conductive heat flux to raise the
core temperature to $T\approx 0.5-0.6T_0$.  However, the action of the
HBI further increases the field line orientation to
$\langle\theta_{\rm B}\rangle> 85^\circ$, and insulates the core which
proceeds to collapse.  The cores in models A5B2 and AR9B2 collapse in
380 and 560 dynamical times after the beginning of the simulation,
respectively, compared with the collapse on a timescale of
$>1000t_{\rm dyn}$ for model T2B2.
\begin{figure}[t]
\includegraphics[width=0.48\textwidth]{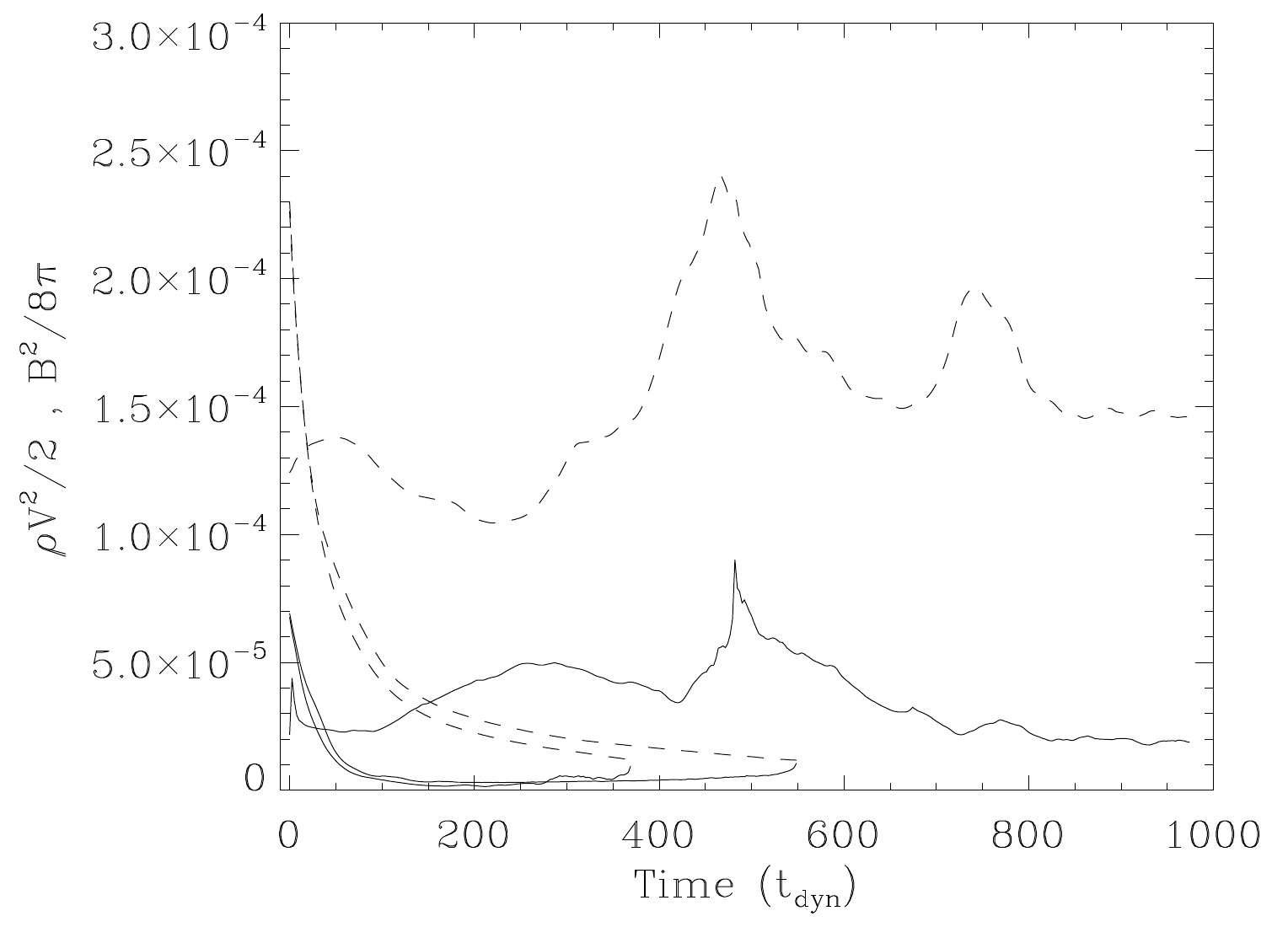}
\caption{Evolution of the kinetic (solid) and magnetic (dashed) energy
density for models T2B2 (ending at 1000$\,t_{\rm dyn}$), A5B2
(380$\,t_{\rm dyn}$), and AR9B2 (560$\,t_{\rm dyn}$). Both components
of energy density are normalized to instantaneous internal energy
density and comprise a small fraction of it.
\label{collapse_energies}} 
\end{figure}

\begin{figure*}[t]
\includegraphics[width=0.5\textwidth]{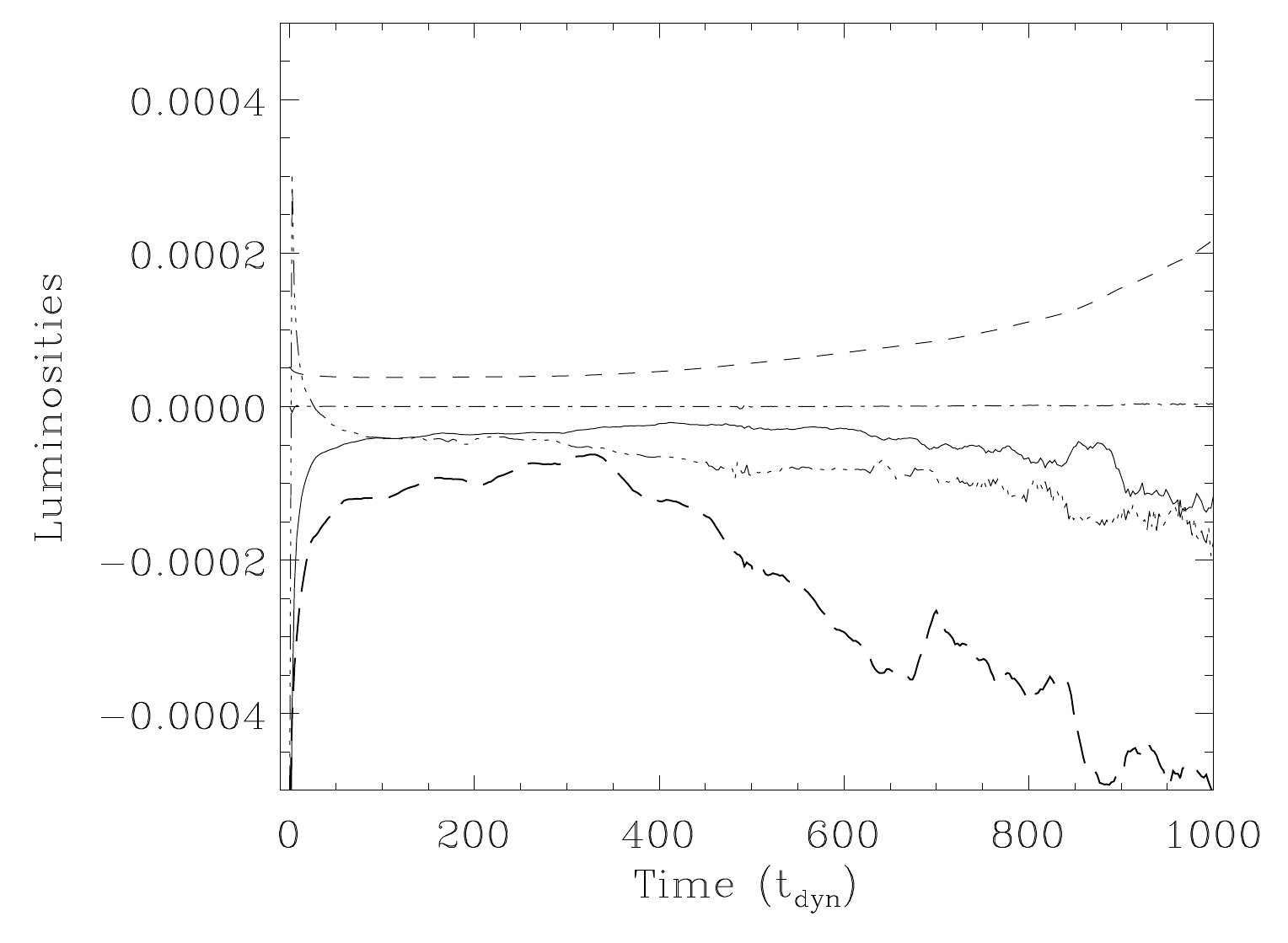} 
\includegraphics[width=0.5\textwidth]{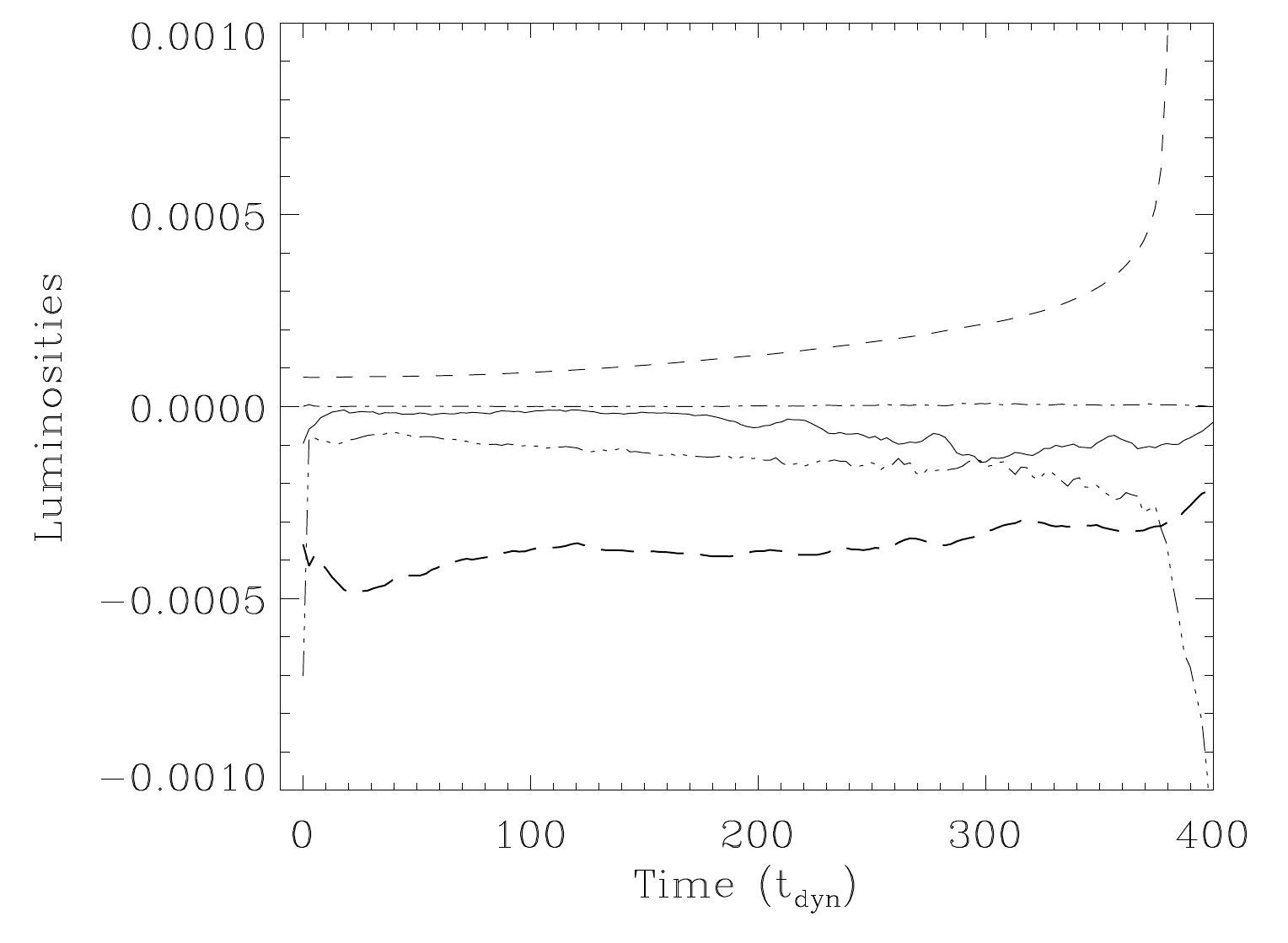}
\caption{Evolution of luminosity components in models T2B2 (left
panel) and A5B2 (right) measured within the sphere of radius of 0.1
from the center of the cluster. Each component is marked with a
different line: anisotropic heat conduction (solid), radiative
(dashed), convective (dash-dot), advective (dash-3dot), and fiducial
value of unbridled heat conduction at the Spitzer value (thick,
long-dashed), all in arbitrary units. The heat flux at full Spitzer
value was divided by a factor of 5 (left panel) and 10 (right) in
order to show it alongside of other components. Negative fluxes are
inflowing and vice versa.
\label{heat_fluxes}} 
\end{figure*}

As was the case for the non-cooling clusters, the modeled cooling
cores remain kinematically quiescent and the magnetic field never
plays a direct dynamical role.  The evolution of kinetic and magnetic
energy density for models T2B2, A5B2, and AR9B2 is shown in
Figure~\ref{collapse_energies}. The kinetic and magnetic energy are a
small fraction of the internal energy of the gas and the core
region. Velocities remain very subsonic, $v<10^{-2}c_s$.  Independent
of the assumed initial geometry of the field lines, HBI does not
result in significant field amplification.

The fact that the cores remain (relatively) kinematically quiescent
results in very little convective heat flux.  To see this,
Fig.\ref{heat_fluxes} shows the evolution of the radial convective,
conductive and advective heat flow for models T2B2 and A5B2 within a
sphere of radius $r=0.1$ along with the total radiative
loses and the fiducial value of  unbridled conductive heat luminosity at 
the Spitzer value, for comparison. To construct these quantities, we start
with the radial components of the advective, convective and conductive
heat fluxes defined by
\begin{eqnarray}
Q_{adv} & = & \frac{\gamma}{\gamma -1} k_{B} \left(\left<n\right> \left<T\right> \left<v_r\right> +
       \left< T \right> \left<\delta n \delta v_r \right> \right) \,, \\
Q_{conv} & = & \frac{\gamma}{\gamma -1} k_{B} \left(\left<v_r\right>
\left<\delta n \delta T\right> + \left<n\right> \left<\delta n \delta
T \right> + \right. \nonumber \\
 & & \left. \left<\delta n \delta T \delta v_r\right>  \right)\,,\\
Q_{cond} & = & -\chi ({\bf\hat{b}\cdot\hat{r}})^2 \,\partial T / \partial r \,\,
\end{eqnarray}
Here, $v_r$ is the radial component of the velocity field.  The
fluctuations ($\delta v_r, \delta n, \delta T$) were defined as
relative to a mean calculated on the surface of the sphere. We then
integrate these components across the surface of the spherical shell
$r=0.1$ to obtain relevant luminosities plotted in
Fig.\ref{heat_fluxes}.  The convective heat flux acts as a cooling
term for the core (Balbus \& Reynolds 2008) but remains significantly
smaller than the other heat fluxes.  As expected, the conductive heat
component initially decreases as the HBI wraps the lines of magnetic
field around the core. At later times ($t > 600\,t_{\rm dyn}$) it
starts increasing again as a consequence of radial field line
stretching in the process of core collapse, which once again
``unlocks" heat conduction along the temperature gradient. Note that
trends in evolution of luminosity components in T and A-models appear
qualitatively similar, however, there is about an order of magnitude
difference in the magnitude of conductive luminosity between the two.
This is a consequence of an early, efficient suppression of
anisotropic heat conduction by magnetic field lines in model A5B2 in
comparison to T2B2. In both runs the field lines eventually saturate
at an angle of $\langle\theta_{\rm B}\rangle\approx 85-88^\circ$, in
such way suppressing the conductive heat flux to a fraction of only
$({\bf \hat{b}\cdot\hat{r}})^2\approx 10^{-3}$ of the Spitzer value
throughout the core volume.  Apart from unbridled heat conduction, the
largest of the remaining four components of luminosity in both models
is the radiative cooling. In terms of luminosity magnitudes, radiative
cooling is followed by advective luminosity which can be interpreted
as heating of the core in the process of adiabatic compression, which
is in turn followed by anisotropic conduction and convective
luminosity.

\begin{deluxetable}{ccccc} 
\tablecaption{Models of Perseus-like cluster. \label{table3}}
\tablewidth{0pt}
\tablecolumns{6}
\tablehead{
\colhead{Model} & 
\colhead{{\bf B}-field} &
\colhead{Resolution} &
\colhead{$\kappa_{\rm aniso}$} &
\colhead{$\Delta t_{\rm collapse}$} \\
\colhead{} & 
\colhead{structure} &
\colhead{(N$^3$)}  &
\colhead{($\kappa_{\rm Spitz}$)}  &
\colhead{(Gyr)} 
}
\startdata
P1 & A    & 100 & 1 & 9      \\ 
P2 & AR & 100 & 1 & 10.5  \\
P3 & A    & 200 & 1 &  8      \\
P4 & A    & 64   & 1 &  6   \\
P5 & A    & 100 & 0 & 2    \\
\enddata
\end{deluxetable}

As stated at the beginning of this Section, the parameter space is
occupied by three distinct groups of systems which end up in either
catastrophic collapse ($t_{\rm cool}/t_{\rm cond}\lesssim25$),
isothermal state ($t_{\rm cool}/t_{\rm cond}\gtrsim10^2$), or as
border line cases.  Border line cases represent the transition
population between the collapse and isothermal group of objects and
their final state is determined by the structure of magnetic field.
The behavior of borderline cases is illustrated by comparing models
T3B2 and A6B2 (which differ only in their initial field structure),
where the former evolves to isothermal state in $80\,t_{\rm dyn}$ and
the latter collapses after $510\,t_{\rm dyn}$.  Another example is
provided by models T3B4, A7B4, and AR11B4, of which only the core in
the T-model evolves towards isothermal state while those in A- and
AR-models collapse. Cores in models A7TN and AR11TN are also border
line cases given that both evolve on very long time scales and that
they seem to hang on a very edge between collapse and isothermality.
Interestingly, the core in A7TN initially evolves towards collapse but
after $\sim 300\,t_{\rm dyn}$ the temperature curve reverses towards
isothermal. The core temperature in AR11TN reminds relatively flat
until late time in the simulation and its final state undetermined.
The cause of different outcomes in these two models is probably
stochastic, driven by a slightly different evolution of the magnetic
field and it underlines the role of the field structure in these
transition population of objects.  The values of $t_{\rm cool}/t_{\rm
cond}$ for TN, B2, and B4 border line cases are 26, 38 and 75,
respectively.  We also find that cores with $t_{\rm cool}/t_{\rm cond}
< 25$ evolve towards collapse regardless of their initial field
geometry (T2, A6, and AR10 in case of both B3 and B4 runs but also
T2B2 and A5B2), while those with $t_{\rm cool}/t_{\rm cond}\gtrsim
250$ all evolve towards isothermal state (T3B3, A7B3, and AR11B3).

\subsection{Models of physical clusters\label{S_perseus}}

In order to offer a more intuitive interpretation of the role of HBI
for a portion of the parameter space populated by real clusters, we
present a second group of calculations where we scale simulation
parameters specifically to match the cooling core of a rich galaxy
cluster.  We mark the position of these models with letter ``P'' in
$(t_{\rm cool}, t_{\rm cond})$ space in Figure~\ref{param_space}. It
is worth noting that characteristic timescales of real cooling core
clusters occupy the range $1 \lesssim t_{\rm cool}/ t_{\rm dyn}
\lesssim 10$ and ${\rm few}\times 0.1 \lesssim t_{\rm cond}/ t_{\rm
dyn} \lesssim 10$, which places them in the collapse region of
Figure~\ref{param_space}.

\begin{figure*}[t]
\center{
\includegraphics[width=0.45\textwidth]{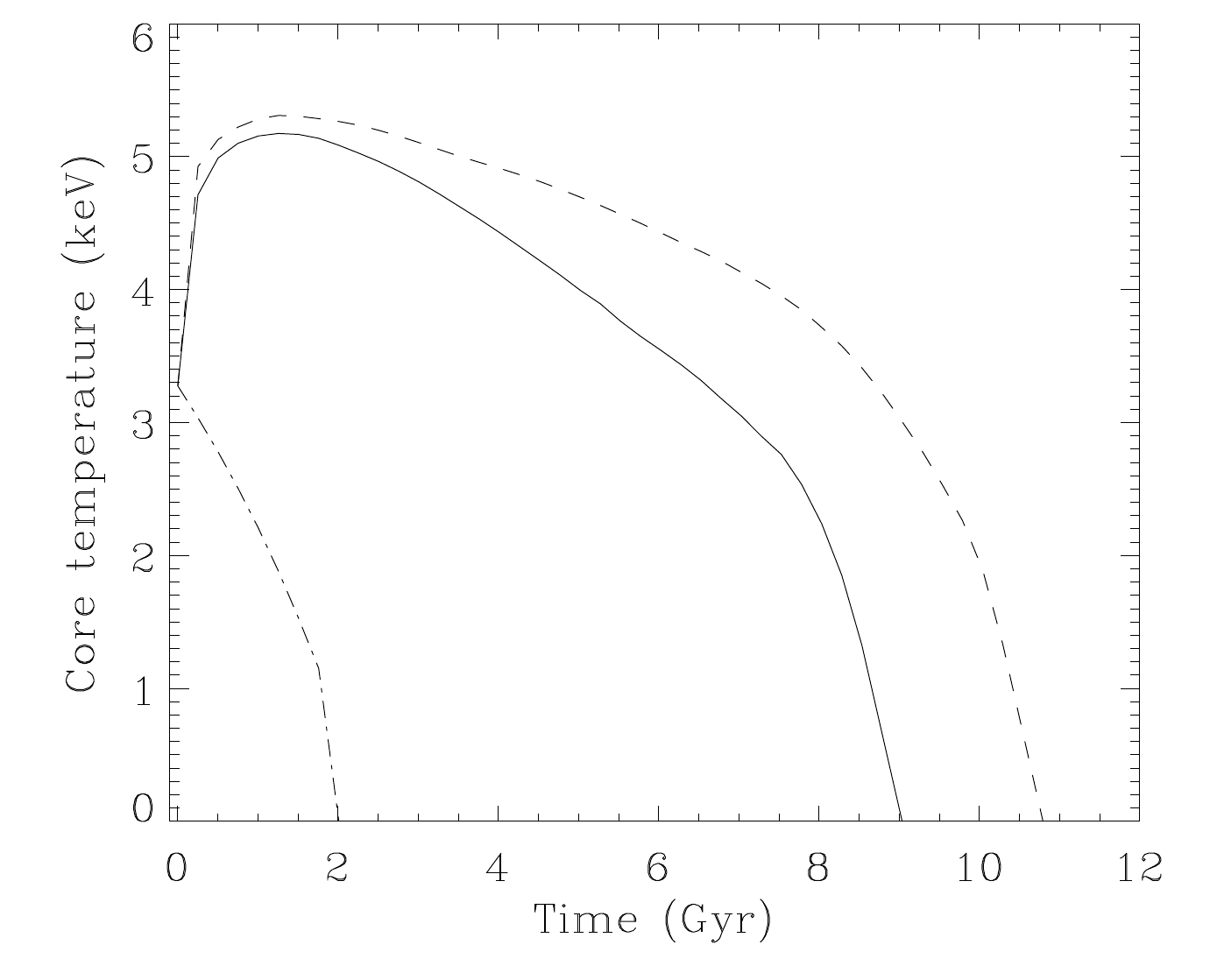} 
\includegraphics[width=0.45\textwidth]{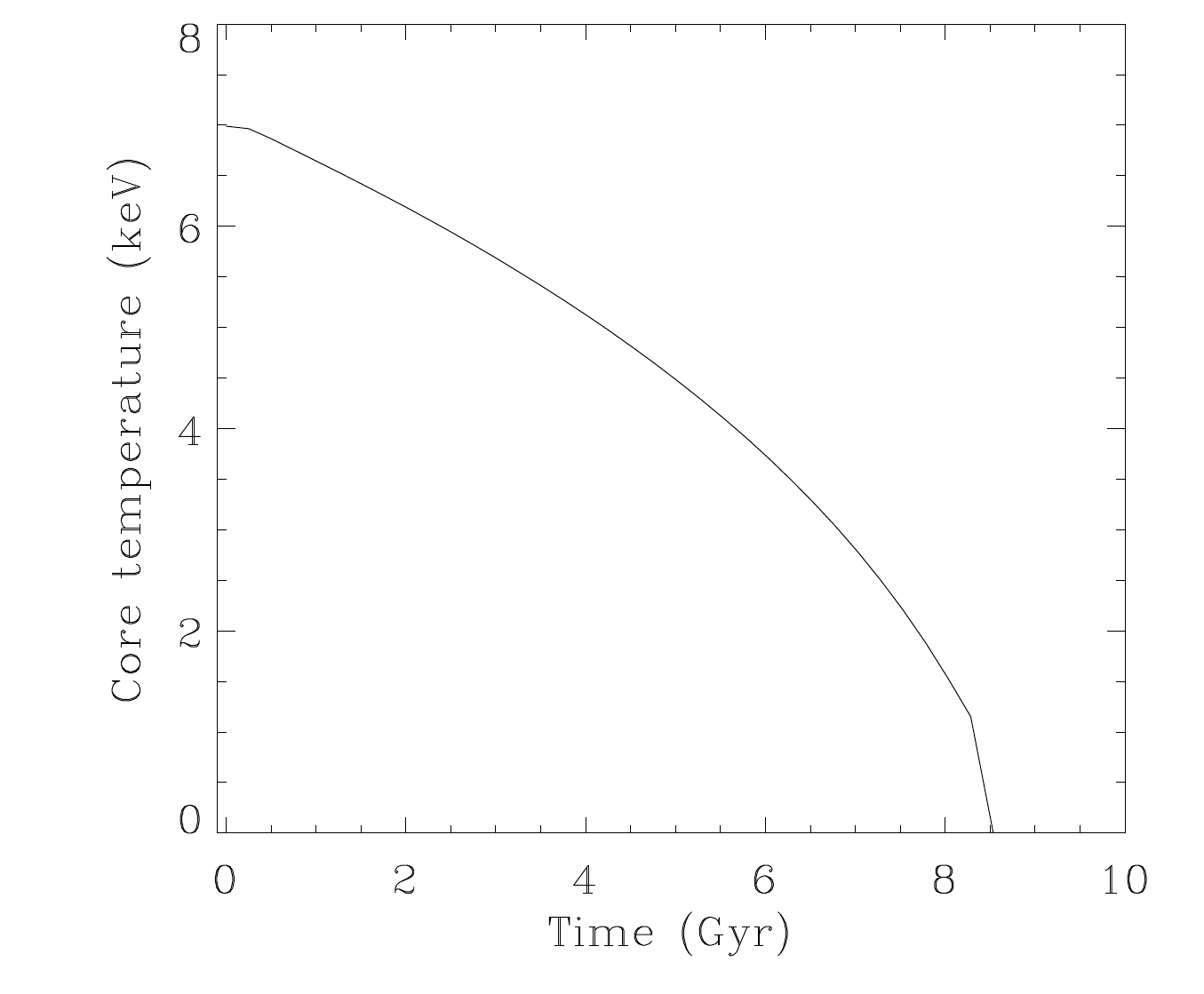}}
\caption{{\it Left:} Evolution of temperature in the core of a cluster
for models P1 (solid), P2 (dashed), and P5 (dash-dot) models. {\it
Right:} Same for P5 only, starting from an isothermal core at virial
temperature of 7~keV.
\label{perseus_temp}} 
\end{figure*}

\begin{figure*}[t]
\center{
\includegraphics[width=0.45\textwidth]{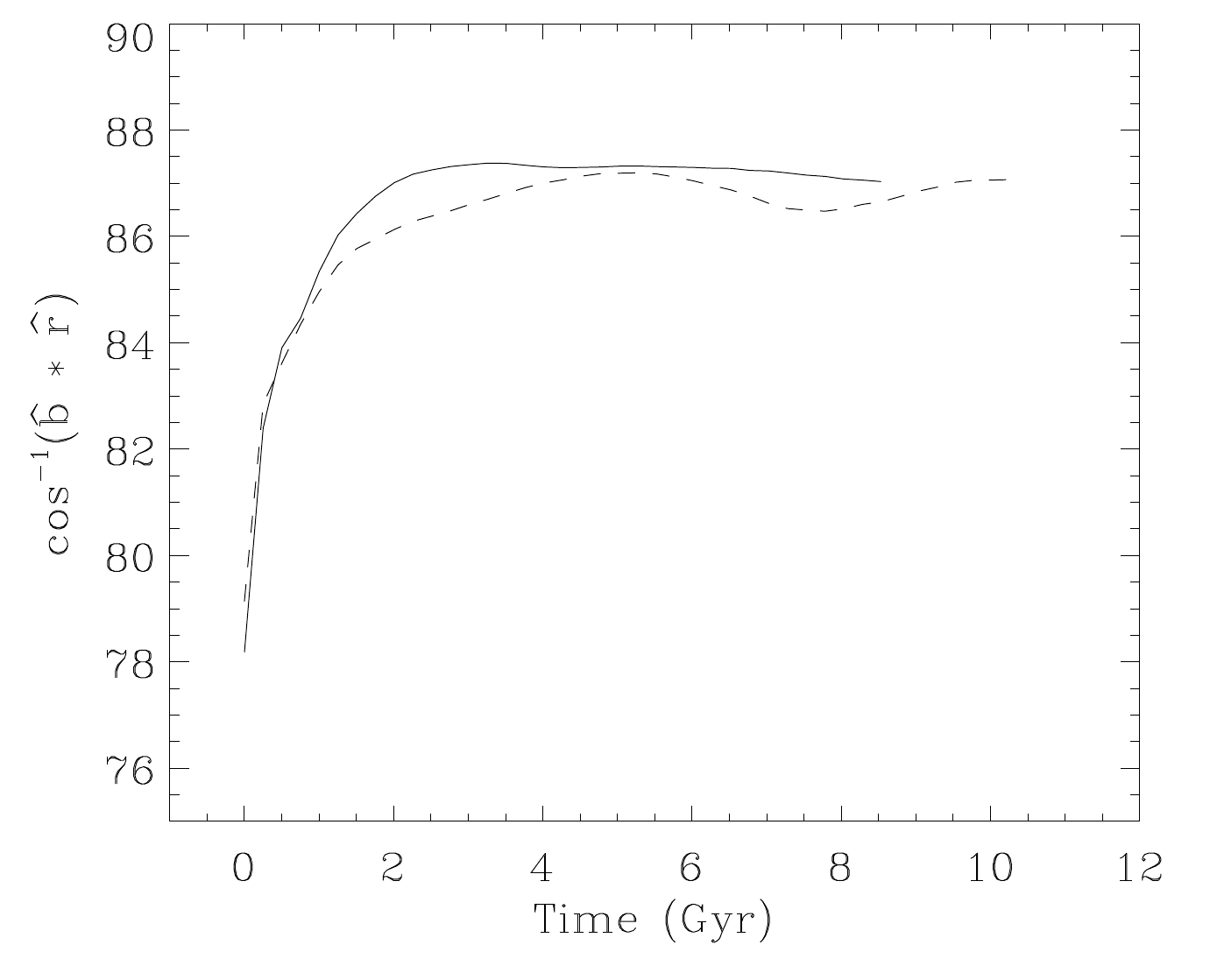} 
\includegraphics[width=0.45\textwidth]{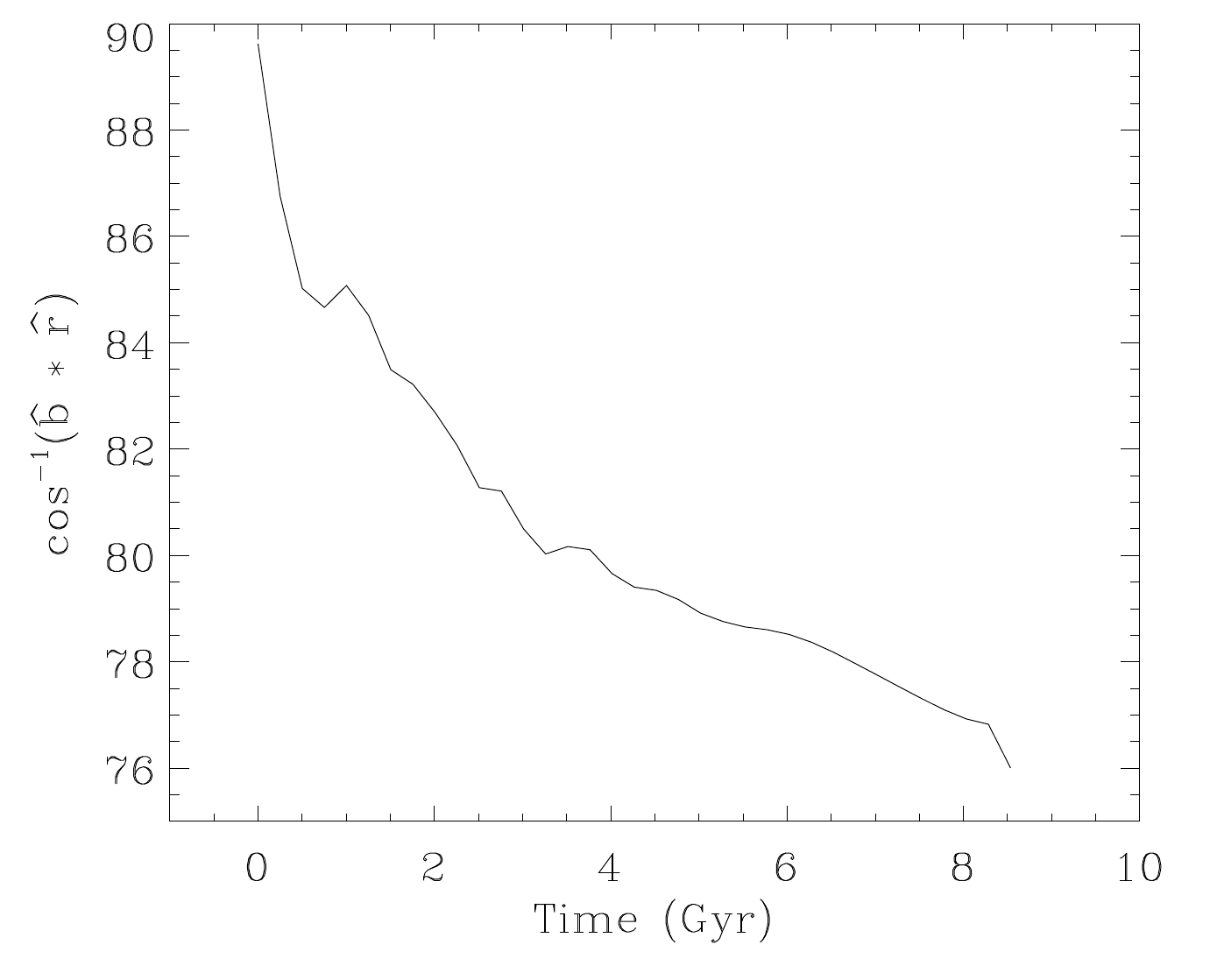}}
\caption{{\it Left:} Evolution of $\langle\theta_B\rangle$ in P1
(solid) and P2 (dashed) models in the central 60 kpc region. {\it
Right:} Same, except for P4 model (pure radiative collapse). Spherical
collapse in absence of HBI leads to smaller $\langle\theta_B\rangle$
and does not maintain the azimuthal orientation of the magnetic field
lines.
\label{perseus_thetab}} 
\end{figure*}

Specifically, we consider a cluster characterized by a virial
temperature of $kT_0=7$\,keV, initial core temperature of $kT_c\approx
3$\,keV, and core radius $R_0 = 100$\,kpc; these properties are
reminiscent of the Perseus cluster.  The gas number density in the
modeled core is $n_0 = 0.03\,{\rm cm^{-3}}$ (somewhat lower than in
Perseus) and the speed of sound is $c_s\approx 900~{\rm km\,s^{-1}}$
in the core center. The core cools radiatively according to the
Tozzi-Norman cooling function \citep{tn01}. For this cluster, we ran a
set of models spanning different magnetic field configurations and
simulation resolutions, as detailed in Table~3.  In this group of
models we test the evolution of the cores with predominantly azimuthal
magnetic field geometry, a scenario that allows us to calculate a
lower limit on the life time of conducting cores.  With the exception
of model-P5, the (anisotropic) heat conduction is set to the Spitzer
value (equation~\ref{eq5}) and mediated by magnetic field in all
models. Model-P5 is a pure radiative collapse model ($\kappa_{\rm
aniso}=0$) given for comparison. The (initial) characteristic time
scales for this system are $t_{\rm dyn} \approx 1.3\times10^8$~yr,
$t_{\rm cond}\approx 4.1\times10^7$~yr, and $t_{\rm cool}\approx
2$~Gyr.  We assume initial values of plasma parameter and the strength
of radial component of magnetic field to be $\beta =10^3$, $B_r = 0$
and $B_r =3\mu G$, for A- and AR-models, respectively.

The collapse of conducting cores is postponed by a factor of
$\sim2-10$ with respect to the time scale for pure radiative collapse.
The cores in models P1 and P2 evolve through an initial heating of the
core during which time the HBI re-orients the field lines in order to
reduce the conductive heat flux.  Figures~\ref{perseus_temp} and
\ref{perseus_thetab} show the evolution of the core temperature and
$\langle\theta_B\rangle$ in P1- and P2-models along with the pure
radiative collapse case, for comparison. The core with initial
temperature of $\sim 3$~keV, in absence of heat conduction heads
towards collapse in only $t_{\rm cool}\approx2$~Gyr (model P5). The
cores in P1 and P2 models evolve through the initial rise in
temperature, have a phase of uniform evolution approximately linear
with time, and in the final 2~Gyr dominated by radiative cooling
behave similarly to the pure radiative collapse case. As remarked in
the previous section, the initial rise in the core temperature in
models with heat conduction is a consequence of non-equilibrium
initial conditions. The intermediate phase of steady, quasi-linear
decay of the core temperature is specific to conducting cores and is
absent in case of pure collapse model.  The presence of HBI is
indicated by the high value of $\langle\theta_B\rangle = 87^\circ$
shown in the left panel of Figure~\ref{perseus_thetab}, contrary to
the pure radiative collapse scenario P5 (right panel) which does not
maintain the azimuthal structure of the field.  The field line
re-orientation appears to saturate at $\langle\theta_B\rangle\approx
87^\circ$ and, from then on, the core undergoes a gradual decline in
core temperature until it reaches a temperature of $kT\sim 2-3$\,keV.
After that time, the core temperature decreases more rapidly due to
the onset of line-cooling within the TN cooling function and the
cooling catastrophe is reached $\sim2$\,Gyr later.  Indeed, this final
stage in the conducting core models is very similar to the pure
radiative collapse case, conduction having very little effect.

If real cluster cores are similar to the cores modeled here, they are
likely to be observed precisely during the long gradual decline phase
(although, compared with model clusters during this phase, real
clusters have a larger fractional temperature decrease as one heads
into the core).  In model-P1, this phase starts from 1~Gyr after the
beginning of the simulation (at which time the core has a temperature
of 5\,keV) and the core collapses 8~Gyr later.  In model P2, the life
time of the core is extended due to the presence of a split monopole,
with collapse occurring 9.5\,Gyr after the start of this decline
phase.  A pure cooling simulation starting from a temperature of
5\,keV undergoes the cooling catastrophe in approximately 4\,Gyr.
Thus, even starting off with rather azimuthal field geometries,
thermal conduction delays the thermal collapse of the cluster core by
at least a factor of 2.

It follows from the findings in previous section that the longest
lived cooling cores are those with large scale magnetic fields
(represented by T-models). In order to estimate the factor by which
core collapse is postponed in these models, we consult our grid of
models outlined in Table~\ref{table1} and \ref{table2}. It is possible
to infer that T-model cores collapse on time scales $2-5$ times longer
with respect to the cores with predominantly azimuthal fields
(A-models), as illustrated in examples of model pairs: T2B2 -- A5B2,
T2B3 -- A6B3, and T2B4 -- A6B4. Hence, if T-model cores evolve $2-5$
times longer than A-model cores, and Perseus-like A-model cores evolve
on time scales $\sim2\,t_{\rm cool}$, than the overall evolution of
collapsing cores can be prolonged by a total factor of
$\sim2-10\,t_{\rm cool}$.

\begin{figure}[t]
\includegraphics[width=0.48\textwidth]{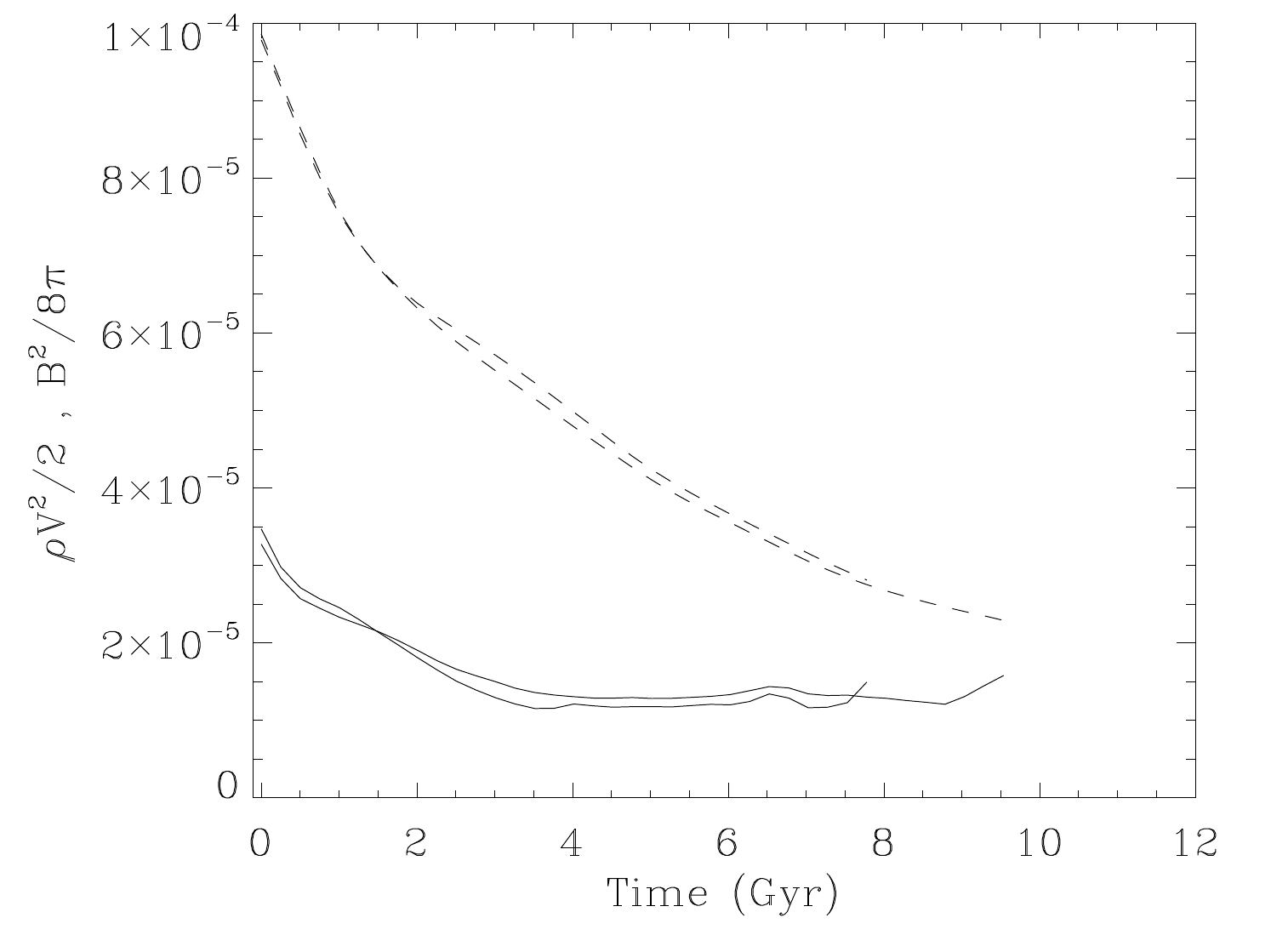}
\caption{Evolution of mean kinetic (solid) and magnetic (dashed)
energy density in P1(shorter) and P2 (longer) normalized to mean
instantaneous energy density.
\label{perseus_energies}} 
\end{figure}

\begin{figure*}[t]
\includegraphics[width=0.5\textwidth]{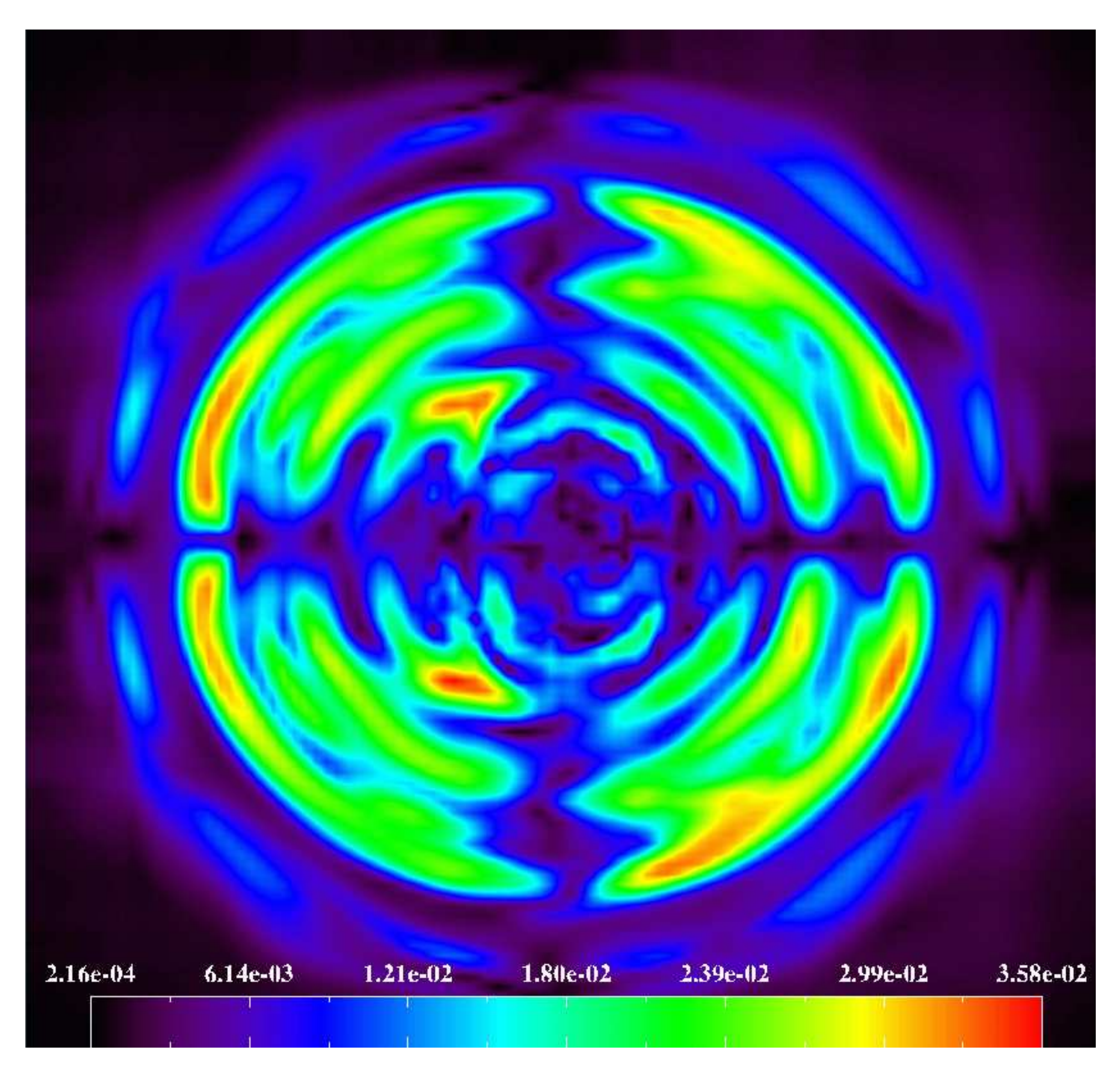} 
\includegraphics[width=0.5\textwidth]{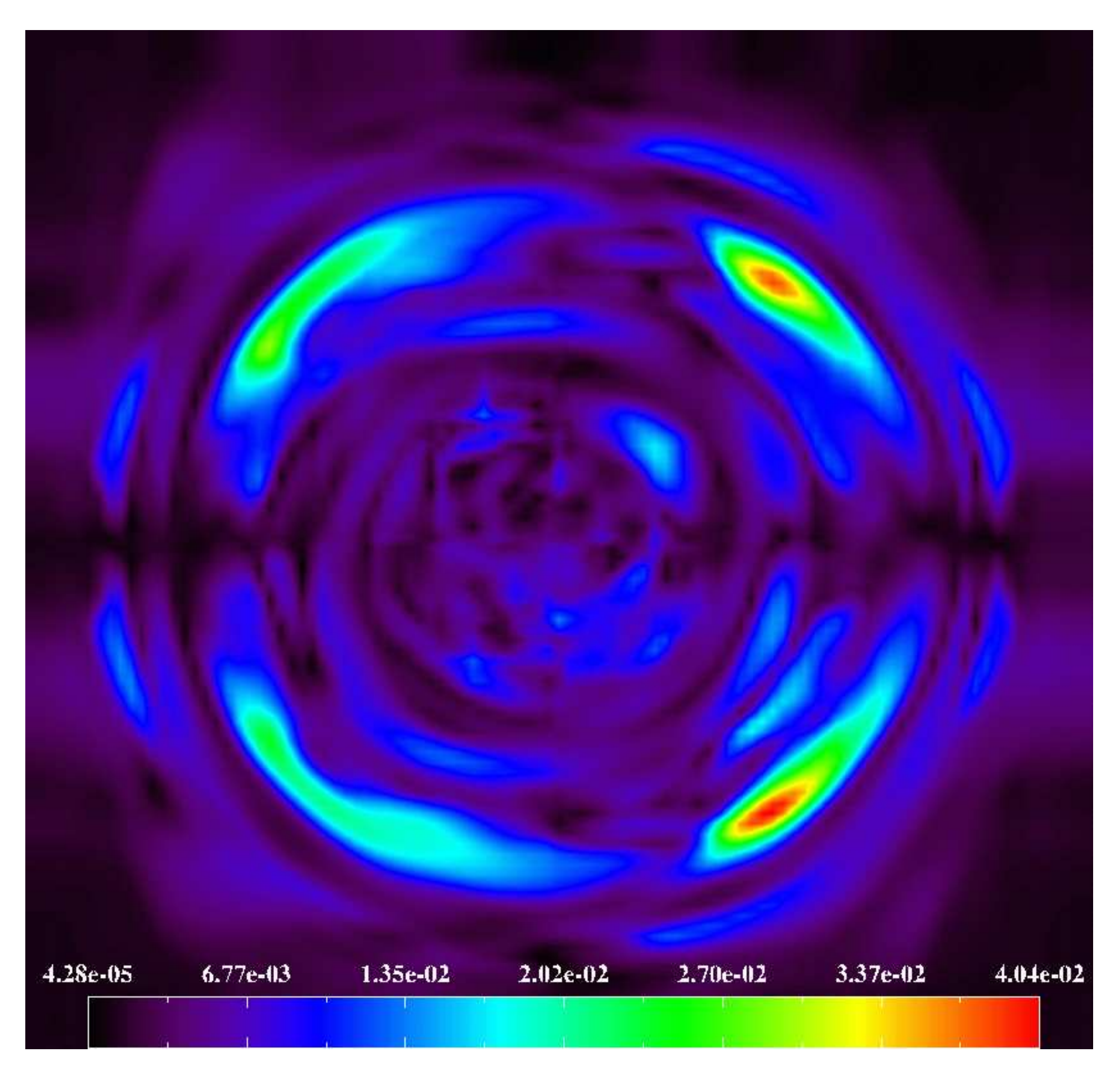}
\caption{Snapshots from P1 model of the Perseus cluster showing the
distribution of velocity magnitude at 500~Myr (left) and 3.5~Gyr
(right) after the beginning of the simulation. Figures show the slice
through the xy-plane of a cluster center. A velocity unit corresponds
to $820\,{\rm km\,s^{-1}}$, implying that highest velocities are of
the order of $30\,{\rm km\,s^{-1}}$. The apparent symmetry in velocity
distribution is a consequence of initial symmetry in magnetic field
geometry.
\label{perseus_velocities}} 
\end{figure*}

\begin{figure}[t]
\includegraphics[width=0.48\textwidth]{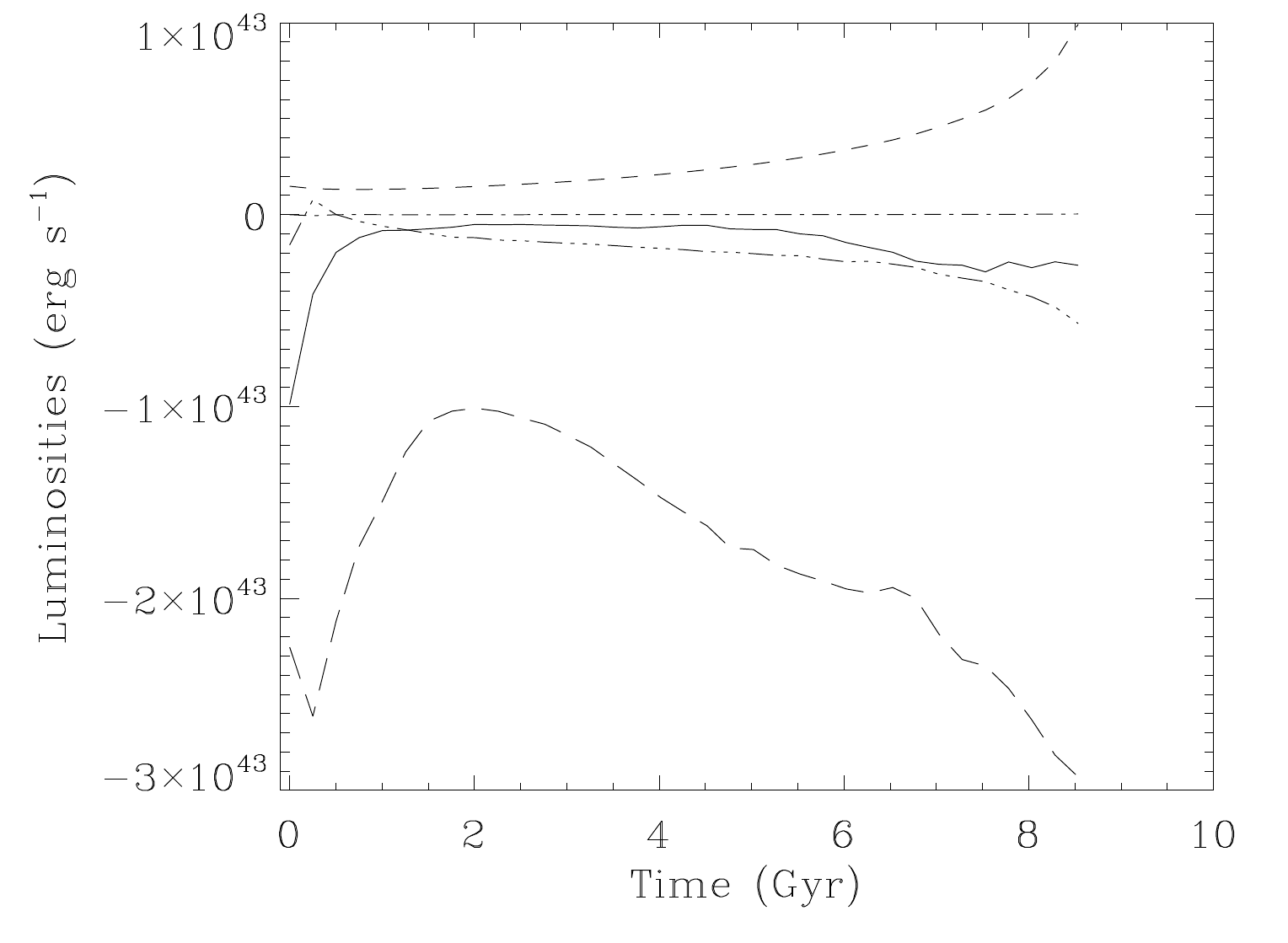}
\caption{Evolution of luminosity components in model P1 measured
within a sphere of radius of 20~kpc. Each component is marked with a
different line: anisotropic heat conduction (solid), radiative
(dashed), convective (dash-dot), advective (dash-3dot), and a fiducial
value of unbridled heat luminosity at the Spitzer value (thick,
long-dashed). The heat flux at the full Spitzer value was divided by a
factor of 10 in order to show it alongside of other
components. Negative fluxes are inflowing and vice versa.
\label{lumin_perseus}} 
\end{figure}

\begin{figure*}[t]
\center{
\includegraphics[width=0.45\textwidth]{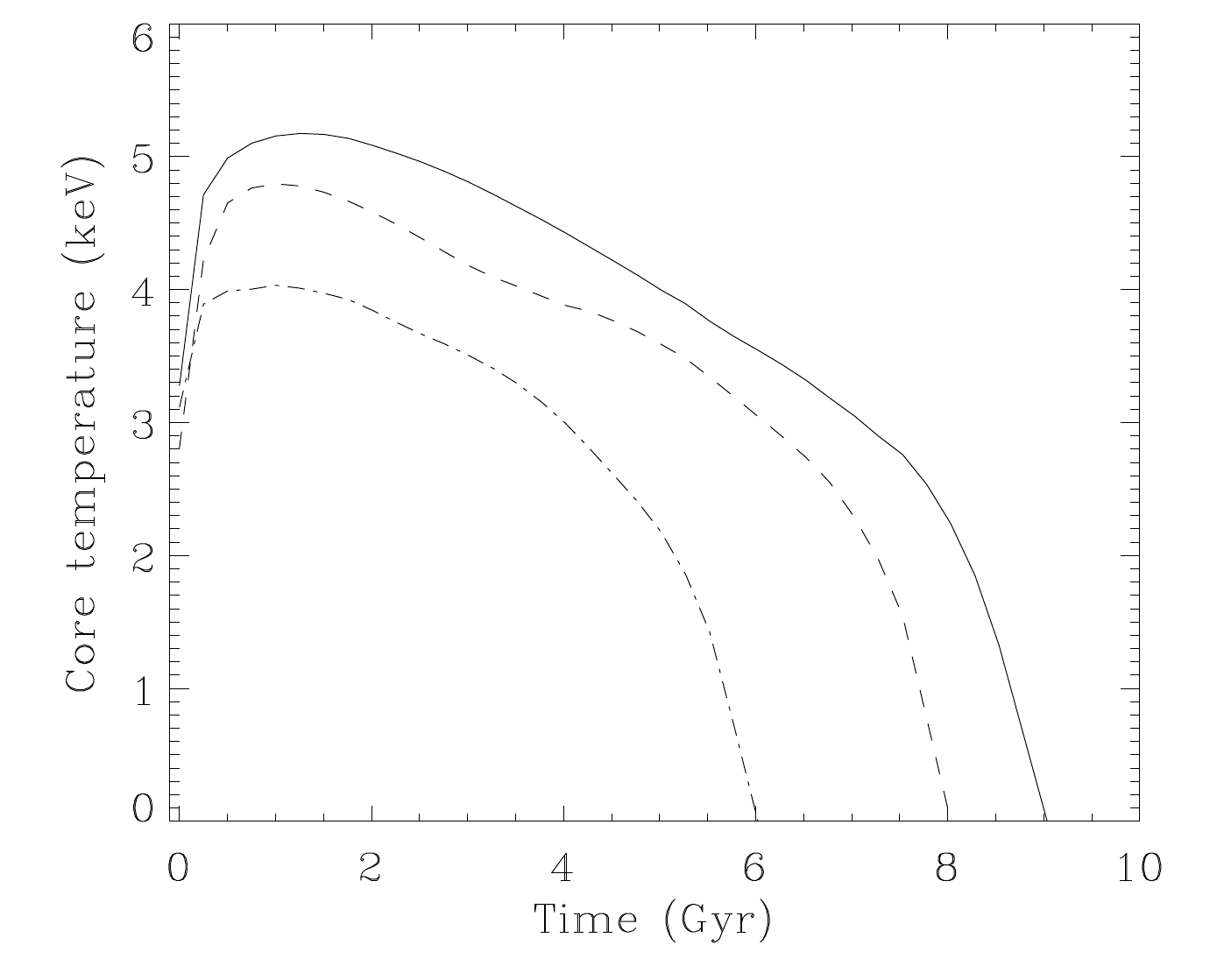} 
\includegraphics[width=0.45\textwidth]{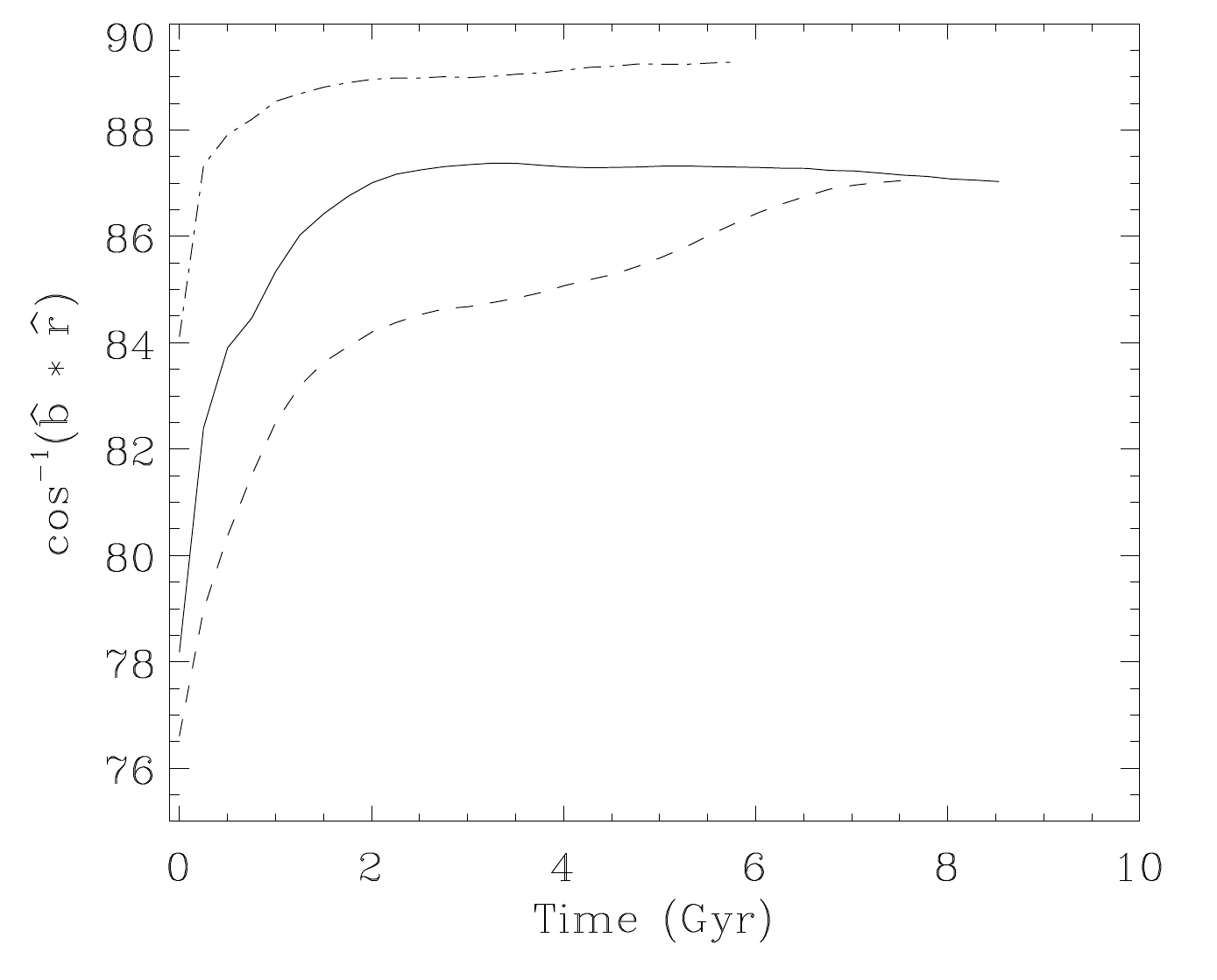}}
\caption{Effect of numerical resolution on evolution of the core
temperature (left) and $\langle\theta_B\rangle$ (right) in P1 (solid),
P3 (dashed), and P4 (dash-dot) runs carried out at resolution of
$100^3$, $200^3$, and $64^3$, respectively.
\label{perseus_resolution}} 
\end{figure*}

As found in Sections~\ref{S_isothermal}--\ref{S_collapse}, these model
clusters cores are characterized by an absence of strong turbulence or
a magnetic dynamo effect.  The magnetic and kinetic energies uniformly
decrease in P1 and P2 models (Figure~\ref{perseus_energies}) and are
only a small fraction of $\sim 10^{-5}-10^{-4}$ of the internal
energy.  Figure~\ref{perseus_velocities} shows the distribution of
velocity magnitude in the xy-plane containing the core center. The
highest velocities in the core are of the order of $30\,{\rm
km\,s^{-1}}$ and are visibly associated with the magnetic field
structure. As the relative strength of magnetic field decreases in the
process of core collapse, the core center becomes kinematically more
quiescent with respect to the initial velocity distribution.

Modeled Perseus-like cores exhibit the strongest heat conduction and
heat flux buoyancy instability within the radius of $\sim20$~kpc -- a
distance comparable to the half-depth radius of the core where the
temperature gradient is steepest. Figure~\ref{lumin_perseus} shows the
evolution of five luminosity components in units of ${\rm
erg\,s^{-1}}$ calculated for a sphere of radius of 20~kpc in model
P1. Note that during the first $\sim 1$~Gyr of evolution, anisotropic
heat conduction overcompensates for the radiative losses and that this
phase is coincident with the initial rapid rise of the core
temperature seen in Figure~\ref{perseus_temp}. Between 1 and 7.5~Gyr
radiative losses are comparable to but larger than the rate of heat
conduction and in this period core temperature steadily declines. In
the remaining period of evolution, the rate of radiative cooling
becomes $\gtrsim 2$ times larger than the anisotropic heat conduction
and the core evolves towards collapse in the fashion similar to a pure
radiative collapse scenario. The rate of energy transported by
convection is consistently $\sim 2$ orders of magnitude lower than
that of the anisotropic conduction and it acts as a cooling term
throughout the simulation. Also included in Figure~\ref{lumin_perseus}
is the fiducial rate of unbridled heat conduction at the Spitzer value
(thick, long dash line), plotted for comparison -- because this
component of luminosity is significantly larger than the others, we
divided it by a factor of 10 in order to show it in the same plot.

Models P1, P3 and P4 constitute a crude numerical resolution study for
these models.  In order to characterize the effect of numerical
resolution on our results we choose the run P1 with numerical
resolution of $100^3$ as a baseline model. With tangled azimuthal
field structure and characteristic coherence lengths of $r_1 = 0.1$
and $r_2=0.087$ this run is a good choice for resolution study,
because the resolution effects are potentially more severe than in the
runs with large scale, loosely tangled field. We carried out
calculations equivalent to P1, at higher resolution of $200^3$ (run
P3) and lower resolution of $64^3$ (run P4).  Interestingly, we find
that stepping up the numerical resolution leads to a non-monotonic
evolution of the collapse timescale (Fig.~\ref{perseus_resolution}).
This behavior can be explained by the effect of resolution on magnetic
field orientation (right panel of Fig.~\ref{perseus_resolution}).
Higher resolution captures the finer, radial structure of tangled
field lines in addition to the dominant azimuthal component, resulting
in the lower initial value of $\langle\theta_B\rangle$. This results
in a highest rate of heat conduction in model P3 (0.04, 0.054, and
0.01 of the Spitzer value for P1, P3, and P4, respectively) and a more
rapid reorientation of the field lines due to HBI when compared to
moderate-resolution model P1.  The consequence of rapid HBI evolution
in model core P3 is that it also collapses before the core in P1. On
the other hand, heat conduction in model P4 is initially at such a low
level that its evolution proceeds fairly similar to a pure radiative
collapse and thus, occurs on a shorter time scale that in models P1
and P3.  Progressing from the simulation with $64^3$ resolution
towards the higher end, the discrepancy in overall evolution time of
the core decreases from $\sim 2$~Gyr to $\sim 1$~Gyr, indicating
convergence.  Based on this we estimate that core evolution in most
affected models in our calculations can be prolonged by $\sim 1$~Gyr.

\section{Discussion}\label{S_discussion}

Heat conduction is in principle capable of providing a large fraction
of the energy necessary to prevent the core collapse in clusters of
galaxies --- indeed, the potential importance of conduction has been
discussed ever since the realization that thermal collapse was a
problem for ICM atmospheres (Binney \& Cowie 1981).  However, attempts
to solve the cooling flow problem with thermal conduction face several
challenges.  Because of a steep temperature dependence of Spitzer
conductivity, heat conduction is expected to play a lesser role in low
mass clusters and groups of galaxies, where most of the gas is below
5\,keV \citep{best07,zn03}. Even unbridled Spitzer thermal conduction
cannot offset radiative loses in these low-mass systems.  Hence, heat
conduction cannot be a single solution to the cooling flow problem
across the full range of masses, and additional sources of heat are
needed to stabilize cooling flows in at least these lower mass
systems. Even in hot systems where conduction is potent, previous
studies have highlighted fine-tuning problems \citep{nulsen82,bd88}.
Under the assumption that a tangled magnetic field essentially acts as
a scalar suppression factor for isotropic conduction, and that the
suppression factor $f$ is a fixed parameter, \citet{kn03} show that
$f$ must be fine-tuned in order to obtain a thermal quasi-equilibrium
resembling real cooling core clusters.  Even then, this equilibrium is
unstable with a growth time $\sim2-5$~Gyr.

Our results indicate that presence of the HBI even further undermines
the role of heat conduction because of its tendency to stifle
conduction by rearranging the lines of magnetic field to be orthogonal
to the temperature gradient.  We find that only systems with a large
disparity in characteristic time scales, $t_{\rm cool}/t_{\rm
cond}\gtrsim 10^2$, can avoid the cooling catastrophe by evolving to
isothermal state.  Systems reminiscent of real clusters, with $t_{\rm
cool}/t_{\rm cond}\lesssim 25$ evolve towards thermal collapse on time
scales $\lesssim 10\,t_{\rm cool}$. For many cooling core clusters
(such as Perseus) this is significantly shorter than a Hubble time,
hence rendering heat conduction alone incapable of rescuing the core
from collapse.

That said, we do find that conduction can significantly increase the
time to thermal collapse, i.e., a significant energy is conducted into
the core before the field-line re-orientation by the HBI seals it off
and enables it to collapse.  At this point, we must acknowledge the
missing ingredients in our idealized cluster models.  A quiescent
cluster core can be significantly disturbed by cosmologically-induced
dynamics (e.g. sub-cluster mergers) and/or powerful outbursts from a
central AGN.  Either of these events will disrupt the azimuthal nature
of the field and may partially reset the evolution being described by
our models.  Thus, while thermal conduction seems unable to stabilize
cooling cores alone, it may still be the primary agent heating cluster
cores provided that it is {\it enabled} by subcluster mergers or AGN.
Note that this is a very different role than that normally envisaged
for AGN in cooling cluster clusters; if correct, {\it AGN are stirrers
and not heaters.}

An interesting property of our models is the absence of strong
turbulent motions. With plasma motions at $\sim10-30\,{\rm
km\,s^{-1}}$ modeled collapsing cores are kinematically quiescent in
comparison with realistic cores, which exhibit turbulent velocities in
the range $\sim 100-200\,{\rm km\,s^{-1}}$ over comparable spatial
scales \citep{cw01,hatch06}. This highlights the necessary role of the
central AGN and cluster mergers as stirrers of the intercluster
plasma.

Both observational and theoretical studies of the cooling problem have
established that the mechanism responsible for the stability of
cooling cores has to be gentle, spatially distributed, and physically
reminiscent of diffusion \citep{kn03,fabian03}. This is an intrinsic
property of heat conduction but is more difficult to understand in
pure AGN heating scenarios.  For example, simple shock heating by AGN
jets fails to stabilize cooling flows in pure hydrodynamic simulations
\citep{vr06} unless a high degree of turbulent heat diffusion is
postulated \citep{bsh09}. In addition, \citet{nkv07} find that the
entropy of the gas in cooling cores is surprisingly low -- at 10s of
${\rm keV/cm^2}$ the gas is on a brink of catastrophic collapse and
yet maintained in that state over a large fraction of the Hubble time.
This presents a challenge for any mechanism which would heat the gas
via strong turbulent motions or shocks, as it is expected to result in
a much higher value of entropy than that observed. The implication of
this discovery is that an unusually delicately balanced feedback
mechanism operates in cooling cores and that heat conduction in
combination with additional mechanisms may play an important role.

\begin{figure*}[t]
\includegraphics[width=1.0\textwidth]{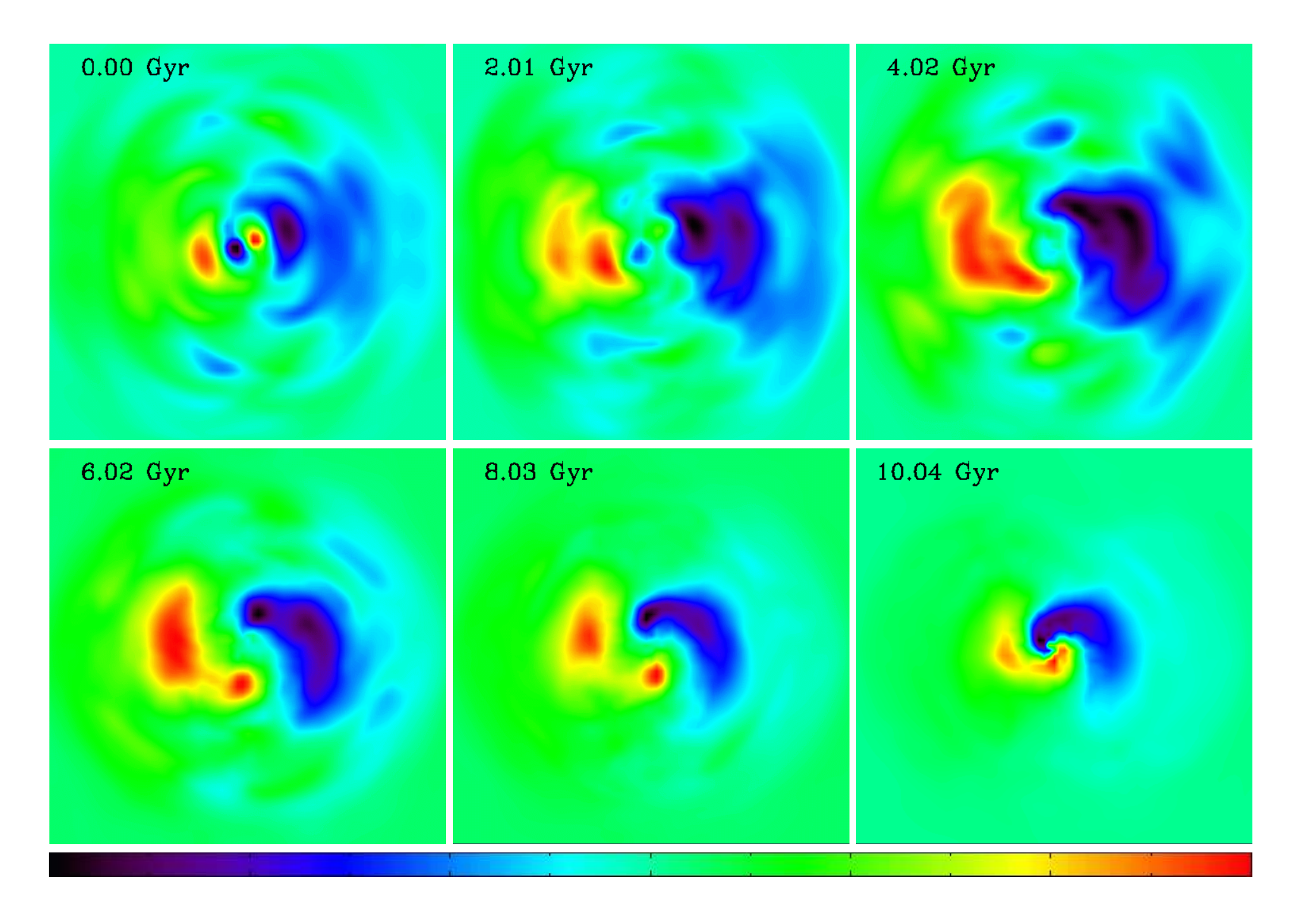}
\caption{Rotation measure maps calculated for model P2 for a line of
sight along x-axis. The colorbar indicates the RM intensity which
ranges from ($-$max, max), where max$ = 150, 110, 95, 120, 170, 265 \;\,
{\rm rad\,m^{-2}}$, for each panel in the time sequence, respectively.}
\label{rm}
\end{figure*}

Another implication of the ubiquity of HBI is that clusters with cool
cores may exhibit a characteristic magnetic field structure, where
field lines are wrapped around the core and have predominantly
azimuthal structure.  Although our model clusters never completely
forget about their initial magnetic field (i.e., T-models and A-models
can be recognized even at late time), the wrapping of field lines into
spherical shells is ubiquitous.  The instability drives the field
lines to almost orthogonal position with respect to the temperature
gradient, $\langle\theta_B\rangle \approx 80-90^\circ$.  This has
several consequences.  Firstly, it creates exactly the kind of
magnetic field geometry envisaged in models of AGN blown bubbles that
invoke ``magnetic draping'' (Ruszkowski et al. 2007).  In these
models, a bubble of relativistic plasma which is buoyantly rising in
the ICM atmosphere is stabilized to Rayleigh-Taylor and
Kelvin-Helmholtz instabilities by a layer of strong magnetic field
that has been swept up on its leading edge.  It is suggested that this
mechanism is responsible for the surprising stability of ``ghost
cavities'' in the ICM of the Perseus cluster, and is an alternative to
models that invoke plasma viscosity (Reynolds et al. 2005).  Secondly,
this characteristic field line geometry leads to the most direct
observable predictions/manifestations of the HBI in the ICM.  Magnetic
fields in the ambient ICM can be probed via the Rotation Measure (RM)
of polarized radio sources in the background.  

The RM is an integral measurement of the effect that magnetic fields
impart on the orientation of the plane of an electromagnetic wave
traveling through the plasma. Hence, an azimuthal field structure
would produce a characteristic radial dependence in the observed RM:
it would vanish at the projected cluster center, and peak at some
projected radius within the cooling core. We calculate this effect for
our Perseus-like models using the expression RM$= 812\,{\rm
rad\,m^{-2}}\int n_e \, {\bf B}\cdot{\bf dl}$, where $n_e$ is the
electron number density in units of ${\rm cm^{-3}}$, ${\bf B}$ is the
vector of magnetic field in units of $\mu {\rm G}$, and {\bf l} is the
depth of the magnetic screen in kiloparsecs. Figure~\ref{rm} shows an
illustrative calculation of the RM for model-P2.  We find that in
Perseus-like models the magnetic field with mean strength of $\langle
B \rangle \sim 0.1 \mu{\rm G}$ produces RM of order $\sim 100 \,{\rm
rad\,m^{-2}}$. The HBI-wrapped field also produces a characteristic
spatial structure to the RM map.  Future observatories with the radio
instruments such as {\it Square Kilometer Array} should provide the
sensitivity needed to identify and study background sources with
sufficient areal density to map out these RM patterns. Note however
that our calculation accounts for the RM arising from the core of a
cluster but does not take into account the effect of magnetic
structures and plasma in the outer parts of a cluster. Therefore, if
the cluster field strength is dominated by the the magnetic fields in
the core, this effect may be measurable.

\section{Conclusions}\label{S_conclusions}

We have performed simulations of cooling cores in clusters of galaxies
with the aim to study the role of anisotropic heat conduction and
recently discovered HBI on the thermodynamic evolution of cores.  Our
models focus on the base state of cluster cores and do not take into
account physics of the central AGN or dark matter or ICM substructure
resulting from hierarchical merging of subcluster units.  We explore
the parameter space of cooling and conduction timescales as well as
different magnetic field configurations.  We summarize our most
important results here:

\begin{itemize}

\item The parameter space of modeled systems can be divided into three
distinct groups of models where collapsing cores occupy the parameter
range $t_{\rm cool}/t_{\rm cond}\lesssim 25$ and isothermal cores
exhibit $t_{\rm cool}/t_{\rm cond}\gtrsim 10^2$.  The range between
these two groups is occupied by a class of systems which final state
is determined by the configuration of magnetic field lines.

\item The efficiency of heat conduction along the lines of magnetic field
ranges between $\sim 10^{-3} - 0.2$ of the Spitzer value, depending on
the initial structure of magnetic field and evolutionary stage of the
core.

\item Modeled conducting cores that correspond to real clusters,
exhibit evolution towards core collapse slower by a factor of $\sim
2-10$ with respect to the time scale for a pure radiative
collapse. For many cooling core clusters (such as Perseus) this is
significantly shorter than a Hubble time, implying that heat
conduction alone cannot rescue the core from collapse. The extent to
which core collapse is postponed in our models is a function of the
initial magnetic field topology, where systems with higher values of
$t_{\rm cool}/t_{\rm cond}$ ratio and field configurations amenable to
conduction result in longer collapse time scales.

\item Magnetic field lines in cores in which HBI is actively operating
rearrange themselves in such way that final value of
$\langle\theta_{\rm B}\rangle \approx 80^\circ - 90^\circ$. In
contrast, cores that are not characterized by HBI have
$\langle\theta_{\rm B}\rangle \approx 60^\circ - 70^\circ$. If seen in
observations, the alignment of field lines across the global
temperature gradient would provide a strong evidence for existence of
MHD instabilities in clusters, which have so far only been considered
theoretically.

\item Heat flux buoyancy instability operating in modeled cores is not
characterized by a strong turbulence and results in kinematically
quiescent cores.  A related consequence of this property is that MHD
turbulence driven by the heat flux instability does not drive a
convective thermal flux sufficient to regulate the core collapse. This
finding is interesting in light of observational and theoretical
studies of the cooling problem which have established that the
mechanism responsible for the stability of cooling cores has to be
gentle and spatially distributed in order to agree with the low levels
of entropy observed in some cooling cores and implies that heat
conduction in combination with additional mechanisms may play an
important role for the cooling problem.

\end{itemize}

The prospects for future work are numerous and they stem from the need
to design more realistic models of cooling cores and also to
understand the interplay of MHD instabilities with other mechanisms
operating in clusters. Future studies will be needed to address the
evolution of cool core clusters within the cosmological framework and
consider the effects of minor and major mergers on gas mixing,
kinematically and MHD driven turbulence, field line tangling, AGN
fueling, and the overall thermodynamic state of a cluster. On the
other hand they will also need to disentangle the effects of plasma
transport processes and magnetic fields via simulations that
incorporate full MHD as well as viscosity, thermal conduction, and
cosmic ray diffusion.

Note that in parallel to this work \citet{pqs09} carried out an
independent study of cool cluster cores with HBI. \citet{pqs09} 
focus on a portion of the parameter space occupied by real 
clusters and consider the role of central entropy
and AGN heating for the final thermodynamical state of a cluster
core. Given different approaches, the two studies complement each
other and in the areas of overlap arrive to similar conclusions.

\acknowledgments

We thank Sean O'Neill and Eve Ostriker for useful and insightful
discussions. Numerical simulations of MHD instabilities in clusters
were carried out on {\it Deepthought}, the University of Maryland high
performance computing cluster and on {\it jxb} at the \'Ecole Normale
Sup\'erieure. TB thanks the UMCP-Astronomy Center for Theory and
Computation Prize Fellowship program for support.  CSR acknowledges
support from the Chandra Theory and Modeling Program under grant
TM7-8009X.

%%%%%%%%%%%%%%%%%%%%%%%%%%%%%%%%%%%%%%%%%%%%%%%%%%%%%%%%%%%%%%%%%%%%%
%%%%%%% R E F E R E N C E S
%%%%%%%%%%%%%%%%%%%%%%%%%%%%%%%%%%%%%%%%%%%%%%%%%%%%%%%%%%%%%%%%%%%%%

\end{document}